\documentclass[%
reprint,
prb,
]{revtex4-2}
\usepackage{graphicx}% Include figure files
\usepackage{dcolumn}% Align table columns on decimal point
\usepackage{bm}% bold math
\begin{document}
\title{Energy symmetry and interlayer wave function ratio of tunneling electrons in partially overlapped graphene}
% Force line breaks with \\

\author{Ryo Tamura}
\affiliation{
Faculty of Engineering, Shizuoka University, 3-5-1 Johoku, Hamamatsu 432-8561, Japan}

\date{\today}% It is always \today, today,

\begin{abstract}
While the exponential decay of tunneling probability with barrier thickness is well known, the accompanying oscillations with thickness have been comparatively less explored.
Using a tight binding model, we investigate an AB-stacked bilayer graphene region acting as an energy barrier between two monolayer graphene leads, under a vertical electric field.
We discuss the case where the energy gap induced by the vertical electric field is comparable to the interlayer transfer integral.
In the up (down) junction, the left and right monolayer leads are connected to different layers (a common layer) of the central bilayer, while the remaining, unconnected layers form armchair-type open edges.
We reveal a characteristic relation between the tunneling probability and the wave function structure. Among the valley-resolved transmission probabilities, only the valley-reversed transmission in the up junction exhibits even symmetry with respect to energy $E$.
This result is counterintuitive. The interlayer wave function ratio $\beta$ is asymmetric in $E$, i.e.,  $\beta(-E) \neq \beta(E)$, and 
electrons cannot bypass the interlayer path in the up junction, whereas they can in the down junction.
We attribute this unexpected symmetry to a self-cancellation effect of $\beta$,  which arises from chiral and rotational symmetry operations combined with the conservation of probability.
Our results demonstrate that the energy dependence of conductance in double junction structures serves as evidence of this effect.
\end{abstract}

\maketitle
%\tableofcontents

\section{Introduction}
\begin{figure}
\begin{center}
\includegraphics[width=0.9\linewidth]{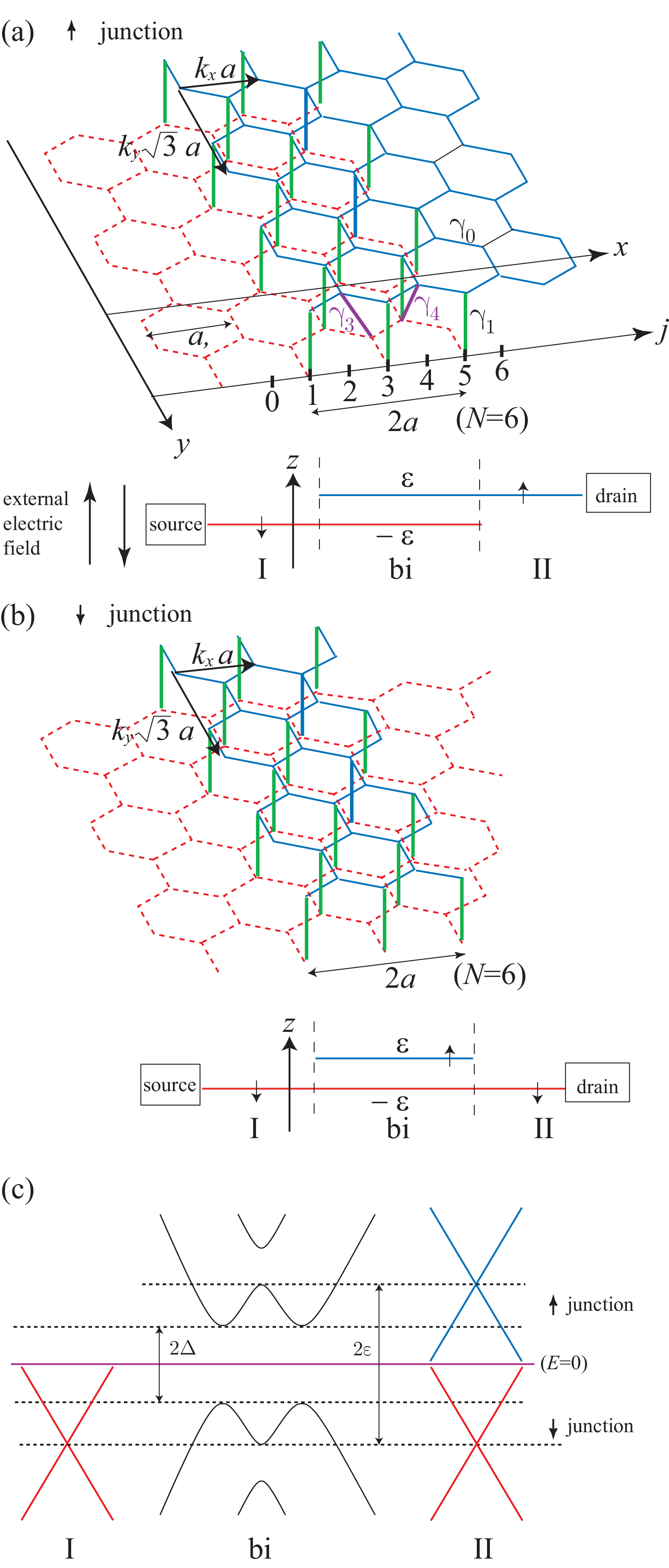}
\caption{
(a) $\uparrow$ and (b) $\downarrow$ junctions.
Each of the expressions $k_x a$ and $k_y \sqrt{3} a$ is accompanied by an arrow indicating the corresponding translational vector.
Regions I and II represent the left and right monolayer regions, respectively, while 
bi denotes the central bilayer region.
The length of the overlapping region is expressed as $(N-2)\frac{a}{2}$, where $N$ is an integer, and $a$ is the lattice constant. As an example, the case of  $N=6$ is illustrated.
 A vertical electric field applied via the top and bottom gate electrodes induces an interlayer site energy difference of $2\varepsilon$.
(c) Dispersion curves in regions I, bi, and II at $k_y=0$.
In region bi, an energy gap with a width of $2\Delta$  emerges due to the interlayer site energy difference $2\varepsilon$, while regions I and II remain gapless.
The $K$-point energy in region II is shifted by $2\varepsilon$ relative to region I in the $\uparrow$ junction, but it remains identical in the $\downarrow$ junction.
}
\label{Fig-intro}
\end{center}
\end{figure}

 Tunneling electrons do not possess a real momentum and manifest themselves in nuclear fusion \cite{1}, scanning tunneling microscopes \cite{2}, tunnel diodes \cite{3}, and tunnel magnetic resistance (TMR) \cite{4}. 
In the barrier region, the component $k_x$ of the wave vector normal to the barrier consists of not only the imaginary part $k_x^{\rm im}$, which determines the decay length but also the real part $k_x^{\rm re}$ that causes the tunneling probability to oscillate with the barrier thickness. 
This oscillatory behavior in tunneling probability has been observed in the TMR \cite{4-2} and theoretically predicted in the side-contacted armchair nanotubes \cite{5, 6}.
 However, the $\uparrow$ junction shown in Fig.~\ref{Fig-intro} (a) and the $\downarrow$ junction shown in Fig.~\ref{Fig-intro} (b) 
provide a more systematic platform for investigating the dependence on barrier thickness and height. We collectively refer to these junctions as partially overlapped graphene (POG) structure, which can be regarded as a series connection of step junctions discussed in Refs \cite{15-step,17-step,po-G-tuika-2-step}.
In these configurations, the source electrode is connected to the bottom ($\downarrow$) layer, while the drain electrode is connected to the top ($\uparrow$) layer in the $\uparrow$ junction and to the bottom ($\downarrow$) layer in the $\downarrow$ junction. Armchair edges appear at the boundaries between monolayer and bilayer regions. We consider the most stable AB stacking configuration. As illustrated in Fig.~\ref{Fig-intro} (c), 
 a vertical electric field opens an energy gap only in the bilayer region, with a width of $2\Delta$ \cite{24,25,85,91}. 
Dual-gate technology enables precise control over the barrier height $\Delta$ and the Fermi level $E$, 
 where $\Delta$ serves as the barrier height \cite{26,27,30,31,29}. 
The bilayer region length -- which corresponds to the barrier thickness --  can be precisely defined using advanced layer-alignment techniques \cite{28}. Although theoretical studies have investigated local gate-induced energy barriers in a pristine bilayer, it remains experimentally challenging to determine the barrier thickness in such structures \cite{32,33}.
As the vertical electric field increases, $\Delta$ increases and eventually saturates \cite{24,25,85,91}. 
This paper considers the case where $\Delta$ is close to its saturation value.

\begin{figure}
\begin{center}
\includegraphics[width=0.7\linewidth]{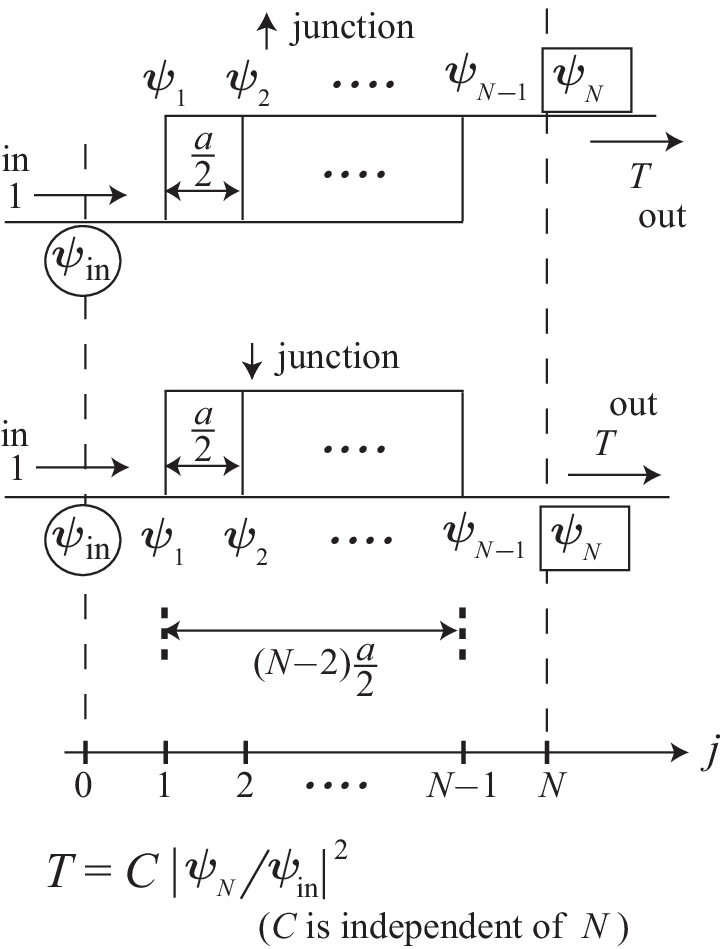}
\caption{Relationship between the wave function and the transmission probability $T$.
The integer $j$ represents the atomic position along the $x$-axis, given by $x = ja/2$, where $a$ is the lattice constant.
The tunneling transmission probability $T$ is approximately proportional to the ratio of the probability densities, $|\psi_N / \psi_{\rm in}|^2$, where $\psi_{\rm in}$ denotes the wave function at site $j = 0$ on the incident ($\downarrow$) layer, and $\psi_N$ is the wave function on the exit layer.
}
\label{New-Fig-2}
\end{center}
\end{figure}

\begin{figure}
\begin{center}
\includegraphics[width=0.7\linewidth]{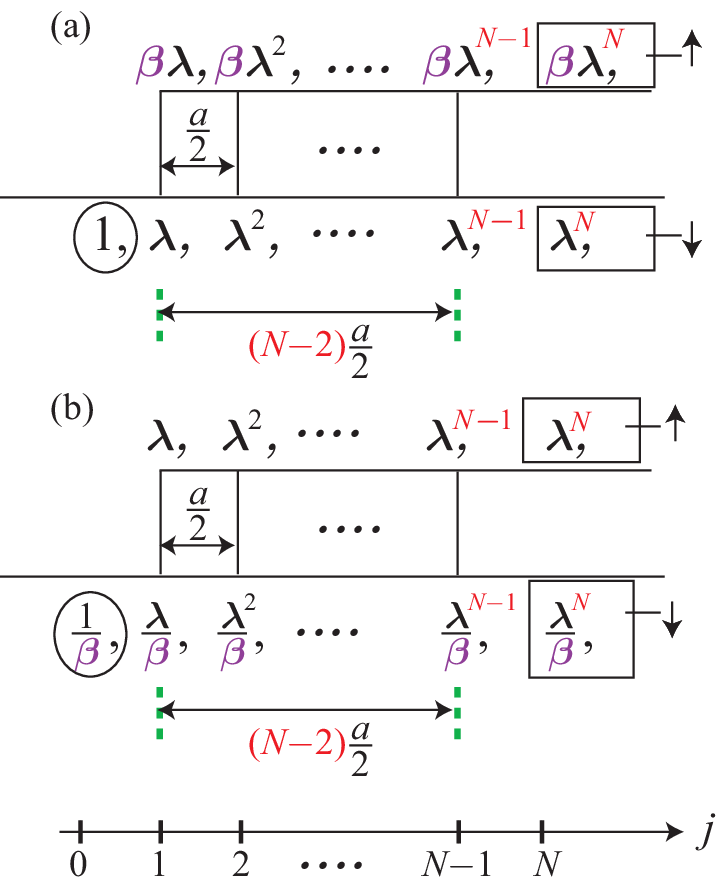}
\caption{ 
Bloch eigenstates of bilayer graphene corresponding to Fig.~\ref{New-Fig-2}.
The parameter $\lambda$ denotes the Bloch factor $\exp(i k_x a / 2)$, and $\beta$ is the ratio of the wave function in the $\uparrow$ layer to that in the $\downarrow$ layer.
(a) $\psi_{\rm in} = 1$; (b) $\psi_{\rm in} = 1/\beta$, where $\psi_{\rm in}$ denotes the wave function at site $j = 0$.
}
\label{New-Fig-3}
\end{center}
\end{figure}

This study focuses on elucidating the connection between the wave function structure and the transmission probability $T$.
As shown in Fig.~\ref{Fig-intro}, we define the coordinate axes such that the atomic $x$ positions are represented by $ja/2$, where $j$ is an integer and $a$ is the lattice constant. The bilayer region is defined by $0<j<N$.
As depicted in Fig.~\ref{New-Fig-2} , let $\psi_{\rm in}$ denote the wave function at site $j=0$ on the incident ($\downarrow$) layer and $\psi_j$ the wave function on the exit layer. The tunneling transmission probability $T$ is approximately proportional to the ratio of probability densities
 $|\psi_N/\psi_{\rm in}|^2$. 
The Bloch eigenstates of bilayer graphene are shown in Fig.~\ref{New-Fig-3} (a) for $\psi_{\rm in}=1$ and in Fig.~\ref{New-Fig-3} (b) for $\psi_{\rm in}=1/\beta$, where $\lambda = \exp(i k_x a / 2)$ is the Bloch factor, and $\beta$ is the ratio of the wave function in the $\uparrow$ layer to that in the $\downarrow$ layer.
 Due to the boundary conditions at $j = 0$ and $j = N$, the states in Fig.~\ref{New-Fig-3} cannot be directly used as $\psi_j$ in Fig.~\ref{New-Fig-2}.
 However, this paper shows that $\psi_j$ is more closely related to Fig.~\ref{New-Fig-3} (b) than to Fig.~\ref{New-Fig-3} (a).
 Although multiple atoms exist at the same $j$, both $\psi_j$ and $\beta$ are consistently defined.
 The $\beta$ factor appears to cancel out in the $\uparrow$ layer in Fig.~\ref{New-Fig-3} (b). This self-cancellation of $\beta$ is analyzed in terms of two key aspects:
 (I) the comparison between the $\uparrow$ and $\downarrow$ junctions, and
 (II) the $N$-dependence of $T$ when the energy $E$ lies within the gap.
To our knowledge, previous theoretical studies on POG structures have not addressed both aspects simultaneously.
References \cite{7-up,12-up,14-up,18-up,21-up,34} considered only the $\uparrow$ junction, while Refs. \cite{9-down,10-down,11-down,22-down,po-G-tuika-down}  focused solely on the $\downarrow$ junction. References \cite{8-updown, 13-updown,16-updown,23-updown,19-updown,20-updown,64} did not investigate the behavior described in (II). The $N$-dependence of $T$  in Refs. \cite{20-updown,64}  does not correspond to the case where 
 $E$ lies within the energy gap. Reference \cite{34} addressed aspect (II) only for the $\uparrow$ junction.

One-dimensional channels are localized at the edges \cite{53,65, 66,67,68} and at the interfaces in the band gap. The interfaces are formed by quadrupolar gate electrodes \cite{69,70,71,72,73,75,gate-1,gate-2},  stacking difference \cite{74,stacking-1,stacking-2,stacking-3,stacking-4}, and the alternating sublattice site energy \cite{76, 77}. The winding number equals the number of the one-dimensional channels if chiral symmetry is present \cite{81, 82}.
The chiral operation flips the sign of the wave function on only one of the sublattices, 
and its generalizations have also been studied \cite{35,36}.
 Although the edge and interface states are irrelevant to the bulk evanescent channels discussed in this paper, the chiral operation remains crucial in the $\beta$ self-cancellation.

In the same way as ballistic electrons, tunneling electrons can be classified based on whether $k^{\rm re}_x$  is near the $K_+$ or $K_-$ corner points in the Brillouin zone -- namely, the $K_+$ or $K_-$  valley -- which are closely related to the valley Hall effect \cite{26,27,30, 37,38,39,40,41}, valley splitter \cite{31,42,43}, valley filter (VF), valley reversal, and optical valley selection \cite{44,45,46,47,48,49,50,51}.
A VF converts a non-valley-polarized current into a current predominantly composed of electrons from one valley \cite{52,53,54,55,56,57,58,59,94}. Valley reversal refers to transmission in which the valley index is flipped during transport \cite{62, 63,64}, and it can coexist with VF functionality \cite{6,34,53,59}.
The charge current is suppressed when two VFs with opposite polarities are connected in series \cite{60,61}. This phenomenon, known as valley blockade (VB), is a probe for the self-cancellation of the interlayer wave function ratio $\beta$.

This paper is organized as follows. Section II presents two types of tight binding models (TB) used in this study: the $\gamma_1\gamma_3\gamma_4$-model and the  $\gamma_1$-model. The simplified $\gamma_1$-model, derived from the $\gamma_1\gamma_3\gamma_4$-model, allows us to analytically determine the wave functions in Sec. III and the transmission probability in Sec. IV.
In Sec. V, by considering chiral operation and $\pi$ rotation in conjunction with the unitarity of the scattering matrix, we prove that the valley-reversed transmission probability in the $\uparrow$ junction is an even energy function.
Section VI compares the analytical results obtained through this approach with the exact numerical results from the $\gamma_1\gamma_3\gamma_4$-model, demonstrating their validity. 
The VB that arises in double $\uparrow$  junctions serves as evidence of the self-cancellation of the interlayer wave function ratio $\beta$.
In Sec. VII, we discuss the relationship between the wave function and transmission probability, as well as the effects of edges and finite size.
In Sec. VIII, we conclude that the POG structure is effective for detecting the phase of the wave function within the energy gap.
Appendix A highlights that the branch cut of the complex square root occurs where the phase is $\pi$.
Appendix B supplements the calculations presented in Section IV.

\section{Tight binding model}
\begin{figure}
\begin{center}
\includegraphics[width=0.7\linewidth]{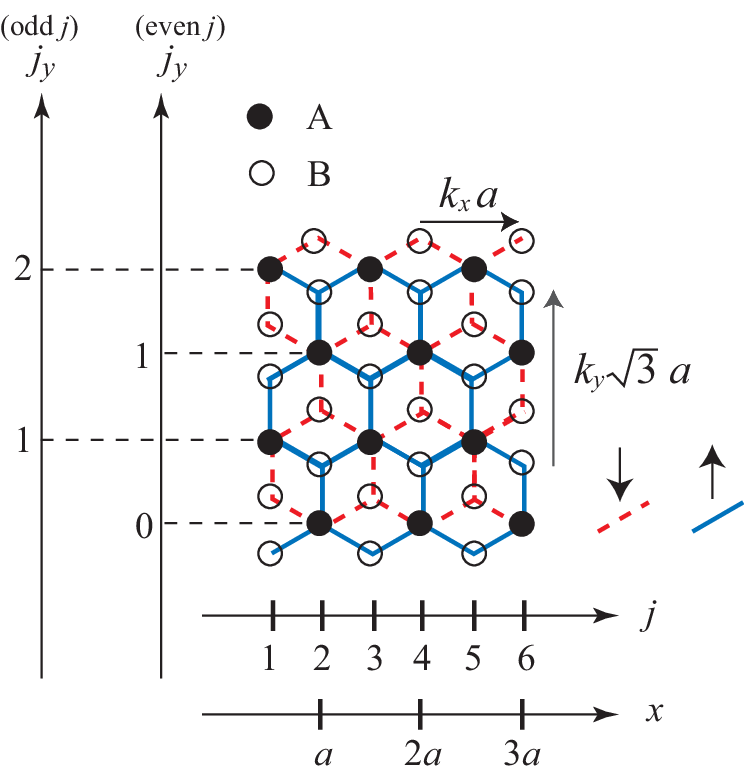}
\caption{
Integers $j$ and $j_y$ are assigned along the $x$-axis and $y$-axis, respectively, to represent atomic positions. Sublattices are labeled A and B, while the layers are labeled $\uparrow$ and $\downarrow$.
The vertical interlayer transfer integral $\gamma_1$ connects the A$_\downarrow$ site and the A$_\uparrow$ site.
Each expression,  $k_x a$ and $k_y \sqrt{3} a$, is accompanied by an arrow indicating the corresponding translational vector.
}
\label{Fig-site}
\end{center}
\end{figure}

Figure \ref{Fig-site} shows how integer labels $(j,j_y)$ and sublattices (A, B) are assigned to the atoms.
When $x=j\frac{a}{2}$ is fixed,  $y_{\rm A}$  increases by $\sqrt{3}a$  for every increment of 1 in $j_y$, 
 and $y_{\rm B}=y_{\rm A} - \circ \frac{a}{\sqrt{3}}$,
 where $a$ denotes the lattice constant, and $\circ$ is a wild card that represents either
 layer symbols $(\downarrow, \uparrow)$ or
  integers $(+1, -1)$.
  One mode corresponds to the other mode as
  \begin{equation}
\circ =\left\{
\begin{array}{c}
 \downarrow = +1\\
 \uparrow = -1
\end{array} 
\right.
\label{notation}
\end{equation}
followed by the interpretation $-\!\downarrow = \uparrow,\;
   -\!\uparrow = \downarrow$.
With $j_y$ fixed, the $y_{\rm A}$ coordinate  is $\frac{\sqrt{3}}{2}a$  smaller for odd $j$ than for even $j$.
These labels allow us to express the wave function as $A_{\circ, j,j_y}$  or $B_{\circ,j,j_y}$.
The TB equation for the bilayer region is represented by 
\begin{eqnarray}
E\vec{d}_{\circ,j} &= &
\left(
\begin{array}{cc}
\gamma_1 & \gamma_4\\
\gamma_4 & \frac{\gamma_3}{\omega^{2\circ}}
\end{array}
\right)
\vec{d}_{-\circ,j} 
+\omega^{-\circ}\left(
\begin{array}{cc}
 0 & \gamma_4\\
\gamma_4 & \gamma_3
\end{array}
\right)
\vec{d}_{-\circ,j}^{\;\;\prime}
\nonumber \\
&&-\left(
\begin{array}{cc}
\circ\varepsilon & 1\\
1 & \circ\varepsilon
\end{array}
\right)
\vec{d}_{\circ,j} 
- \left(
\begin{array}{cc}
 0 & \omega^\circ \\
\omega^{-\circ} & 0
\end{array}
\right)
\vec{d}_{\circ,j}^{\;\;\prime},
\label{TB-eq}
\end{eqnarray}
 where
 $$\vec{d}_{\circ,j}=\left(\sqrt{\omega}\right)^{(-1)^{j+1} }\left(A_{\circ,j,0},  B_{\circ,j,0}\right),$$
 $$\vec{d}_{\circ, j}^{\;\;\prime}=\vec{d}_{\circ, j-1}+\vec{d}_{\circ, j+1},\;\; \omega=e^{i\frac{\sqrt{3}}{2}k_y a},$$ and the intralayer nearest-neighbor transfer integral $\gamma_0$ is negative, with  $\gamma_0 =-|\gamma_0|=-1= -3.12$~eV \cite{84}.
 Unless otherwise noted, we adopt units where $|\gamma_0|=1$.
There are two ways to define sublattices in bilayer graphene. One definition assigns the lattice points such that the vertical interlayer transfer integral $\gamma_1$ occurs between A sites, while the other assigns them such that it occurs between AB sites. The chiral operation transforms  ($A_\circ, B_\circ$)  into ($\circ A_\circ, -\circ B_\circ$)  
under the former definition, and into ($A_\circ, -B_\circ$)  under the latter. 
This paper adopts the former definition since it is suitable for treating the $\uparrow$ and $\downarrow$ layers symmetrically.

We obtain the TB equation for the monolayer regions by setting all the interlayer transfer integrals to zero.
  We assume that the $y$ component of the wave vector $k_y$  is a good quantum number, and  thus $A_{\circ, j,j_y}=\omega^{2j_y}A_{\circ,j,0}$ and $B_{\circ,j,j_y}=\omega^{2j_y}B_{\circ,j,0}$.
 We use two $\pi$ orbital TB models.
 In the $\gamma_1\gamma_3\gamma_4$-model,  $\gamma_1=0.377$~eV, $\gamma_3=0.29$~eV, and $\gamma_4=0.12$~eV \cite{84}. 
 The $\gamma_1$-model is identical to  the $\gamma_1\gamma_3\gamma_4$-model
  except that $\gamma_3=\gamma_4=0$.
   The energy $E$ represents the Fermi level relevant for electron transport.
  The external vertical electric field induces a non-zero $\varepsilon$.
  They are under the control of dual gate electrodes.

 \section{Wave function for a general $k_y$ calculated with the $\gamma_1$-model}
 We designate the right monolayer region with the symbol $\bullet$,  which also follows rule (\ref{notation}), 
	and refer to the junction between the $\bullet$ and $\downarrow$ layers as the $\bullet$ junction.
 When $\gamma_3=\gamma_4=0$, the general solution to the TB equation at the $\bullet$ junction is given by
\begin{equation}
\vec{d}_{\downarrow, j}^{\rm \;I} =\sum_{\nu=\pm}\sum_{\varsigma=\pm}\frac{1}{\sqrt{J_{\nu,\downarrow}}}
\xi^{(\varsigma)}_{\nu, {\rm I}} \tilde{\lambda}_{ \nu,\downarrow}^{\varsigma\nu j}
\left(
\begin{array}{c}
 \tilde{f}_{\nu,\downarrow}
\\
1
\end{array}
\right)
\label{mono-left}
\end{equation}
in the left monolayer region I $(j \le 0)$, 

\begin{equation}
\vec{d}_{\bullet, j}^{\rm \;II} =\sum_{\nu'=\pm}\sum_{\varsigma=\pm} \frac{1}{\sqrt{J_{\nu',\bullet}}}
\xi^{(\varsigma)}_{\nu', {\rm II}} \tilde{\lambda}_{\nu',\bullet}^{\varsigma \nu' (j-N)}
\left(
\begin{array}{c}
 \tilde{f}_{\nu',\bullet}
\\
1
\end{array}
\right)
\label{mono-right}
\end{equation}
in the right monolayer region II $(j \ge N)$, and
\begin{eqnarray}
\left(
\begin{array}{c}
\vec{d}_{\downarrow,j}^{\rm \; bi}
\\
\\
\vec{d}_{\uparrow,j}^{\rm \; bi}
\end{array}
\right)
&=&
\sum_{\tau=\pm}\sum_{l=\pm} \sum_{\varsigma=\pm}  \eta_{\tau,l}^{(\varsigma)} \lambda_{\tau,l}^{\varsigma j}
\left(
\begin{array}{c}
f_{\tau,\downarrow,l}
\\
1
\\
\rho_{\tau,l}\beta_l f_{\tau,\uparrow,l}
\\
\rho_{\tau,l}\beta_l
\end{array}
\right)
\label{bi-wf}
\end{eqnarray}
in the bilayer region $(1\leq j \leq N-1)$, 
 where the overlap length is $(N-2)\frac{a}{2}$.
 The sublattice wave function ratio is
 \begin{equation}
 \tilde{f}_{\nu,\circ}=
\frac{\circ(E+\circ\varepsilon)\omega^\circ}{\nu\frac{\varepsilon}{|\varepsilon|}\sqrt{(E+\circ \varepsilon)^2-s^2 }-i s}
\label{f-mono}
 \end{equation}
 in Eqs.~(\ref{mono-left}) and (\ref{mono-right}), and
  \begin{equation}
 f_{\tau,\circ,l}=
\frac{\circ(E+\circ\varepsilon)\omega^\circ}{\circ\tau\sqrt{p+liq-s^2 }-i s}
\label{f-bi}
 \end{equation}
in Eq.~(\ref{bi-wf})  with notations
\begin{equation}
 p=E^2+\varepsilon^2,\;q=\sqrt{(4\varepsilon^2+\gamma_1^2)(\Delta^2-E^2)} ,
\end{equation}
 and $s=\sin \left( \frac{\sqrt{3}}{2}k_y a\right)$.
 We consider the gap region $|E|< \Delta$,
 where $\Delta=\frac{\gamma_1|\varepsilon|}{\sqrt{4\varepsilon^2+\gamma_1^2}}$ stands for the half energy gap width \cite{24, 25,85}.
 $\xi$ ($\eta$) denotes the mode amplitudes of the monolayer (bilayer) regions.
 The interlayer wavefunction ratio is expressed as the product of
\begin{equation}
\rho_{\tau,l}=\omega^2\frac{\tau\sqrt{p-s^2 +ilq}+is}{\tau\sqrt{p-s^2 +liq}-is}
\end{equation}
and
\begin{equation}
\beta_l=\frac{2\varepsilon E-ilq}{\gamma_1(E-\varepsilon)}.
\label{beta}
\end{equation}
In the limit where the $\rho$ has negligible influence, the $\beta$  effectively represents the interlayer wave function ratio.  
  
  The condition that Eqs.~(\ref{mono-left}) and (\ref{mono-right})  must be in an extended state determines $s_{\rm max}$, 
the maximum allowable $s$ as 
\begin{equation}
s_{\rm max} =\left\{
\begin{array}{ccc}
|E+\varepsilon|  & \cdots &  \mbox{$\downarrow$ junction   }\\
||E|-|\varepsilon||   &  \cdots & \mbox{ $\uparrow$ junction } 
\end{array} 
\right. 
\label{smax}
\end{equation}
Equation (\ref{f-bi}) determines the Bloch factors $\lambda_{\tau,l}$ in Eq.~(\ref{bi-wf})
  as
  \begin{equation}
\lambda_{\tau,l}=\mu+\sqrt{\mu^2-1},
\label{def-lambda}
\end{equation}
 where
 \begin{equation}
\mu=-\frac{\omega^\circ}{2}\left(1+\frac{E+\circ \varepsilon}{f_{\tau,\circ,l} }\right).
\label{mu}
\end{equation}
Equations (\ref{def-lambda}) and (\ref{mu}) equal $\exp(ik_xa/2)$ and $\frac{1}{2}(\lambda^{-1}+\lambda)$, respectively, where $k_x$ denotes the wave vector component with a non-zero imaginary part, indicating its complex nature within the energy gap.
Although $f_{\tau,\circ,l}$ depends on the layer index $\circ$,  $\lambda_{\tau,l}$  does not depend on $\circ$.
Replacing $f$ with $\tilde{f}$ in Eq.(\ref{mu}), we obtain the Bloch phase factor $\tilde{\lambda}$ in Eqs.~(\ref{mono-left}) and (\ref{mono-right}).
 $J_{\pm,\circ}$ denotes ${\rm Re} (\omega^{-\circ} \tilde{f}_{\pm,\circ} ) {\rm Im} (\tilde{\lambda}_{\pm,\circ})$, and the probability flow equals
 $\sum_{\nu}|\xi_\nu^{(+)}|^2-|\xi_\nu^{(-)}|^2$.
 $|\tilde{\lambda}| =1$ for the monolayer region, while  $|\lambda| < 1$ for the bilayer region.
 The superscript $\varsigma$ of $\xi$ ($\eta$) represents the propagation (decay) direction.

  By relating the TB equation to the Dirac equation under the effective mass approximation, $f$ and $\tilde{f}$  represent the orientations of a pseudospin \cite{86,87,88}.
At the interfaces I-bi and bi-II, smaller pseudospin mismatch leads to higher transmission probabilities.
Therefore Eqs.~(\ref{f-mono}) and (\ref{f-bi}) indicate that the bilayer mode $\tau$ is dominant in the transmission under conditions
 \begin{equation}
 \tau=\frac{\varepsilon}{|\varepsilon|}\nu =\bullet \frac{\varepsilon}{|\varepsilon|}\nu', 
\label{dominant-path}
 \end{equation}
  and $(E\pm \varepsilon)^2 \simeq p+liq$. 
  The following identity demonstrates that the latter condition is satisfied when $|\varepsilon| \gg \gamma_1$.
  \begin{equation}
\left|\frac{(E\pm\varepsilon)^2}{p\pm i q}-1\right|^2=\frac{\gamma_1^2}{\varepsilon^2+\gamma_1^2-E^2}
\end{equation}
Reflecting Eq.(\ref{dominant-path}), $T_{\bullet\nu,\nu} \gg T_{\bullet\nu,-\nu}$.
The mode that satisfies $\vec{d}_{j+1} \simeq e^{\pm i \frac{2}{3}\pi} \vec{d}_j$ is designated as the $K_\pm$ valley mode. 
This valley classification correlates with the $x$-axis orientation choice.
 Since $\tilde{\lambda} \simeq e^{i\frac{2}{3}\pi}$,  the $\xi_{\nu, \circ}^{(\varsigma)}$  mode corresponds to the $K_{\varsigma\nu}$-valley mode,  and $\nu$ serves as a valley index.
  We introduce notation $\lambda_\tau \equiv \lambda_{\tau,+}$ and
\begin{equation}
 \lambda_{\tau} =|\lambda_{\tau}|e^{i(\phi_\tau+\tau\frac{2}{3}\pi)}
 \label{lambda-phase}
 \end{equation} 
  where  $\phi_\tau$  signifies the relative phase shift of $\lambda_\tau$ with respect to $\tau\frac{2}{3}\pi$.
  Note that $\lambda_{\tau,-}=\lambda_{\tau}^*$.
 We prove that $|\lambda_{\tau}| < 1$ and $|\phi_\tau| \ll 1$ in Appendix A.
 Equation (\ref{lambda-phase}) illustrates that
  the valley of  mode $\eta^{(\varsigma)}_{\tau,l}$ becomes $K_{\varsigma\tau l}$.   
The degree of freedom $l$ corresponds to complex conjugation and is independent of $\varsigma$.
Complex conjugation correlates with reversing the propagation direction in a propagating wave, but it becomes independent of the decay direction in a decaying wave.

 The valley conservation ($\nu=\nu'$) and valley reversal ($\nu=-\nu'$) manifest themselves in the $\downarrow$ and $\uparrow$ junctions, respectively,
  in Eq.~(\ref{dominant-path}).
 With abbreviation $\eta_1=\eta_{\tau,+}^{(+)}, \eta_2=\eta_{\tau,+}^{(-)}, 
 \eta_3=\eta_{\tau,-}^{(+)}, \eta_4=\eta_{\tau,-}^{(-)}, \lambda=\lambda_{\tau,+}$ and $\beta= \beta_+$, 
 we approximate Eq.~(\ref{bi-wf})  as
\begin{equation}
\left(
\begin{array}{c}
\vec{d}_{\downarrow,j}^{\rm \; bi}
\\
\\
\vec{d}_{\uparrow,j}^{\rm \; bi}
\end{array}
\right)
 \simeq  \left. 
\sum_{\nu=\pm} \left(
\begin{array}{c}
  g_{\downarrow}(j) \left[
 \begin{array}{c}
 \tilde{f}_{\nu,\downarrow}
\\
1
\end{array}
\right]
\\
 \\
  \omega^2 g_{\uparrow} (j) \left[
 \begin{array}{c}
 \tilde{f}_{-\nu,\uparrow}
\\
1
\end{array}
\right]
 \end{array}
 \right)\right|_{\tau=\frac{\varepsilon}{|\varepsilon|}\nu}
 \label{xi-down-up}
\end{equation}

\begin{equation}
g_{\downarrow}(j)= \eta_{1}\lambda^{j}+ \eta_2\lambda^{-j}
+\eta_3(\lambda^*)^{j}+\eta_{4}(\lambda^*)^{-j}
\label{xi-down}
\end{equation}
\begin{equation}
g_{\uparrow}(j)= \beta(\eta_{1}\lambda^{j}+ \eta_2\lambda^{-j})
+\beta^*(\eta_3(\lambda^*)^{j}+\eta_{4}(\lambda^*)^{-j})
\label{xi-up}
\end{equation}
under the conditions $|\varepsilon| \gg \gamma_1$ and $|s| \ll | \sqrt{p-s^2+iq} | $.
When we suppress $\tau$,  $\tau$ is limited to Eq.~(\ref{dominant-path}).

Suppose $N \gg 1$ and the $K_\nu$ electron is incident from region I.
 The wave function (\ref{xi-down-up}) decays
 with $j$ except when $j$ is very close to $N$.
It is equivalent to condition  $|\eta_2|,|\eta_4| \ll |\eta_1|,|\eta_3|$.
Applying this condition to the boundary condition $g_{\uparrow}(0)=0$, we derive
\begin{equation}
\frac{\eta_3}{\eta_1}=-\frac{\beta}{\beta^*}\;.
\label{eta3-eta1}
\end{equation} 
Using Eqs.~(\ref{eta3-eta1}), $ g_{-\bullet}(N)=0$, and 
\begin{equation}
g_\bullet(N) =e^{\nu' i\frac{2}{3}\pi} g_{\bullet}(N-1) ,
\label{connect-N}
\end{equation}
we obtain
\begin{equation}
\frac{\eta_2}{\eta_1}=\frac{\beta}{\beta^*}|\lambda|^{2N}+\delta_{\uparrow,\bullet}
\left(
\frac{\beta}{\beta^*}-1
\right)
\lambda^{2N}
\label{eta2}
\end{equation} 
and
\begin{equation}
\frac{\eta_4}{\eta_1}=-\frac{\beta}{\beta^*}\left[
|\lambda|^{2N}+\delta_{\downarrow,\bullet}
\left(
\frac{\beta}{\beta^*}-1
\right)
(\lambda^*)^{2N}
\right].
\label{eta4}
\end{equation} 
In the derivation of Eqs.~(\ref{eta2}) and (\ref{eta4}), we use approximation $\lambda_\tau \simeq e^{i\tau \frac{2}{3}\pi}$
and $\tilde{\lambda}\simeq e^{ i\frac{2}{3}\pi}$.
However we cannot replace $\lambda^N_\tau$ with   $e^{i\tau \frac{2}{3}\pi N}$,  because $N \gg 1$.
When $j$ is not close to $N$,
\begin{equation}
\left (
|g_\downarrow(j)|, |g_\uparrow(j)
|\right) \simeq
2|\beta\eta_1|
\left(
\left 
|{\rm Im}\left(\beta^{-1}\lambda^j
\right) \right|,
 \left|
{\rm Im}\left( \lambda^j \right)\right|
\right)
\label{xi-j}
\end{equation} holds. At the output interface $j=N$,
\begin{equation}
\left (
|g_\downarrow(N)|, |g_\uparrow(N)
|\right)=
4|{\rm Im} (\beta)\eta_1
|\left (
\left 
|{\rm Re}\left(\beta^{-1}\lambda^N
\right)
\right|, 0
\right)
\label{xi-j-d}
\end{equation} 
for the  $\downarrow$ junction, and
\begin{equation}
\left (
|g_\downarrow(N)|, |g_\uparrow(N)
|\right)=
4|{\rm Im} (\beta)\eta_1
|\left (0,
\left |
{\rm Re}\left( \lambda^N \right)
\right|
\right)
\label{xi-j-u}
\end{equation} 
for the $\uparrow$ junction. 
When Eqs.~(\ref{xi-j}), (\ref{xi-j-d}), and (\ref{xi-j-u}) are compared, it is apparent that the wave function exhibits a phase shift of 
 $\frac{\pi}{2}$ between the entrance and the exit.

\section{Transmission probability for zero-$k_y$ calculated with the $\gamma_1$-model}
We consider the transmission probability $T_{\nu',\nu}$ from valley $K_\nu$ to valley $K_{\nu'}$ at $k_y = 0$. Using the $\gamma_1$ model, analytical expressions have been derived: Ref. \cite{64} obtained $T_{\nu',\nu}$ in the band region $\Delta < |E| < |\varepsilon|$, and Ref. \cite{34} obtained it in the gap region $|E| < \Delta$, given by
\begin{equation}
T_{\nu',\nu}= \left|  \left(\tilde{t}_{\bullet}\Lambda^N
\sum_{n=0}^\infty \left(r_{\downarrow}\Lambda^N r_{\bullet}\Lambda^N\right)^n
t_\downarrow\right)_{\nu',\nu} \right|^2,
\label{multiple-T}
\end{equation}
where $n$ signifies the number of times the wave travels back and forth across the bilayer region, and  $\Lambda=$ diag($\lambda_{+,+},\lambda_{-,+},\lambda_{-,-},\lambda_{+,-})$.
We obtain $\lambda$ for the band region by replacing $iq$ by $|q|$ in Eq.~(\ref{f-bi}).
$t$ ($\tilde{t}$) denotes the transmission matrix at the entrance (exit) interface. 
 Matrixes $r$ represent the reflection.
   Subscript $\downarrow$ ($\bullet$) corresponds to the entrance (exit).
   In the band region, $|\lambda|=1$, and thus, we cannot neglect multiple reflection terms with $ n \geq 1$ in Eq.~(\ref{multiple-T}).
  In the gap region,  however, $|\lambda| < 1$, and Eq.~(\ref{multiple-T}) is approximated by
\begin{equation}
T_{\nu',\nu}= \left|\left(\tilde{t}_{\bullet}\Lambda^N
t_{\downarrow}\right)_{\nu',\nu}\right|^2,
\label{direct-T}
\end{equation}
 when $N \gg 1$. 
 Equation (\ref{direct-T}) can be approximated for the dominant path (\ref{dominant-path})  as follows.
\begin{equation}
T_{\bullet \nu,\nu}=\frac{\gamma_1^6}{q^6}(\varepsilon^2-E^2)^2
\frac{Z_{\bullet \nu,\nu}^2}{|\zeta_{\bullet,+}\zeta_{\bullet,-}\zeta_{\downarrow,+}\zeta_{\downarrow,-}|^2}
\label{T-smat}
\end{equation}
where 
\begin{equation}
Z_{\nu,\nu}=\left. \frac{1}{\sqrt{p}} (|E+\varepsilon| +\sqrt{p})^2{\rm Re}\left(\beta^{-1}\lambda_{\tau}^N\right)
\right|_{\tau=\frac{\varepsilon}{|\varepsilon|}\nu}
\label{Z-dd}
\end{equation} 
for the $\bullet=\downarrow$ junction,
\begin{equation}
Z_{-\nu,\nu}=\left. 2\varepsilon\left(1+\frac{|\varepsilon|}{\sqrt{p}}\right){\rm Re}\left(\lambda_{\tau}^N\right)
\right|_{\tau=\frac{\varepsilon}{|\varepsilon|}\nu}
\label{Z-ud}
\end{equation} 
for the $\bullet=\uparrow$ junction, and
\begin{equation}
\zeta_{\circ,\pm}=1+\frac{(E+\circ\varepsilon)^{\pm 1}}{iq}{\rm Re} \left( \frac{\circ 2\varepsilon E + iq}{(\sqrt{p+iq})^{\pm 1}}\right).
\label{zeta}
\end{equation} 
Appendix B shows the derivation of Eq.~(\ref{T-smat}).
The Re factors in Eqs.~(\ref{Z-dd}) and (\ref{Z-ud}) are represented by
\begin{equation}
{\rm Re}\left(\lambda_\tau^N\beta^{-\delta_{\bullet,\downarrow}}\right)
=|\lambda_\tau|^N|\beta|^{-\delta_{\bullet,\downarrow}}\cos\theta_{\tau,\bullet},
\label{Re-beta}
\end{equation}
where 
\begin{equation}
\theta_{\tau,\bullet}=\tau\frac{2}{3}\pi N+\phi_\tau N-\varphi\delta_{\bullet,\downarrow},
\label{Re-beta-theta}
\end{equation}
$\varphi$ denotes the phase of $\beta$, and $\tau=\frac{\varepsilon}{|\varepsilon|}\nu$.

\section{energy symmetry for the $\uparrow$ junction}
\begin{figure}
\begin{center}
\includegraphics[width=0.7\linewidth]{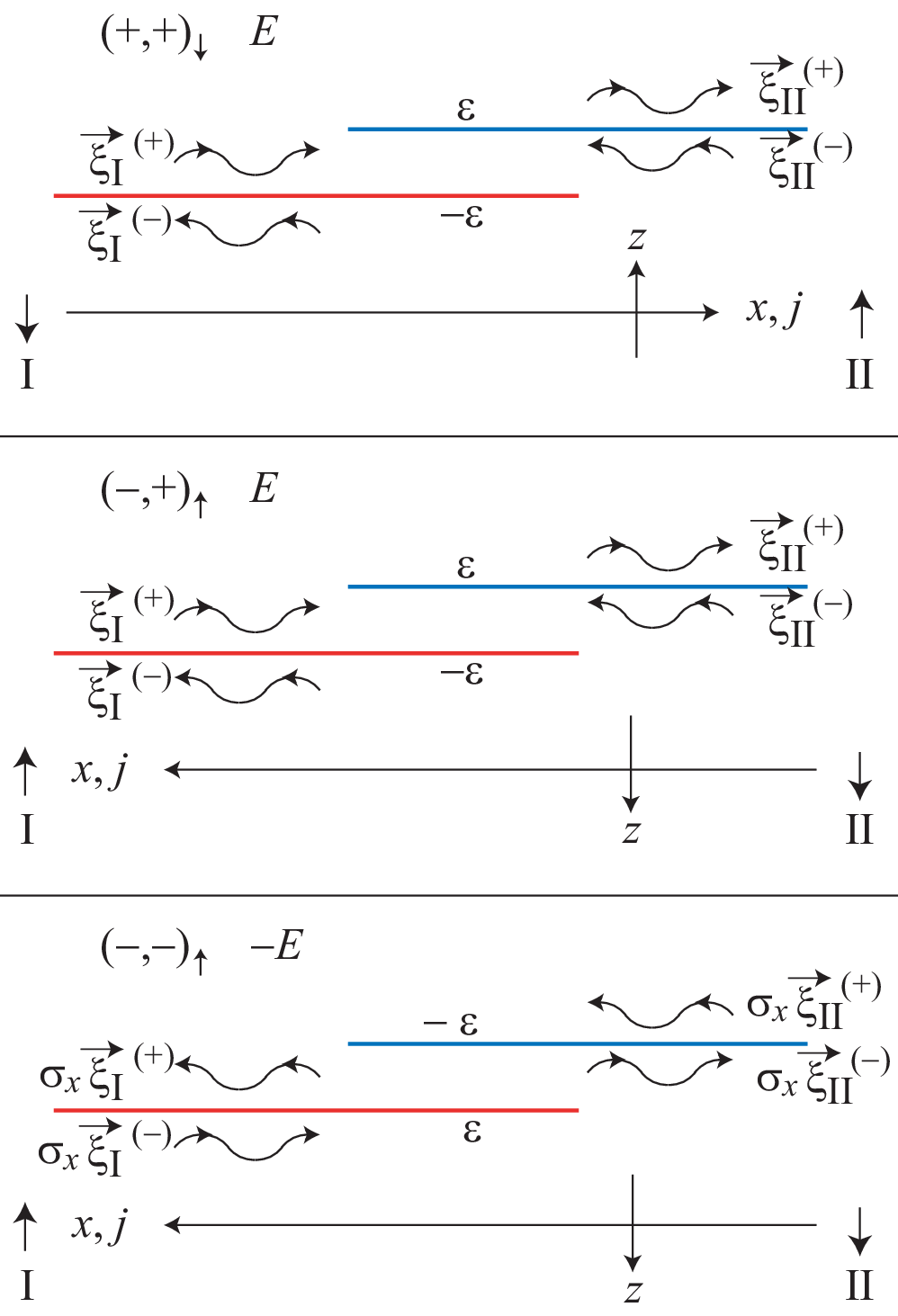}
\caption{Operations representing symmetry.
The meanings of the symbols $(+,+)_\downarrow$, $(-,+)_\uparrow$, and $(-,-)_\uparrow$  are explained in the main text.
The original $\uparrow$ junction $(+,+)_\downarrow$ is transformed into 
 $(-,+)_\uparrow$ by a $\pi$ rotation about the $y$-axis
 The direction from $\downarrow$  to $\uparrow$  aligns with the positive direction of the $x$-axis.
Under the chiral operation, accompanied by the inversion of the signs of  $E$ and $\varepsilon$, $(-,+)_\uparrow$ changes into $(-,-)_\uparrow$.
The arrows of wavy lines indicate the direction of the probability flow.
}
\label{Fig-sym}
\end{center}
\end{figure}

This section proves the energy symmetry of the intervalley transmission probabilities for the $\uparrow$ junction.
This symmetry  exactly holds  in the $\gamma_1$-model, and
Eqs.~(\ref{mono-left}) and (\ref{mono-right}) can be employed for this proof. In this section, however, Eqs.~(\ref{mono-left}) and (\ref{mono-right})  are approximated as
\begin{equation}
\vec{d}_{\circ, j}^{\rm \;I} =\sum_{\nu=\pm} 
\left(
\xi^{(+)}_{\nu, {\rm I}} \Omega^{\iota_1\nu j}+
\xi^{(-)}_{\nu, {\rm I}} \Omega^{-\iota_1\nu j}
\right)
\left(
\begin{array}{c}
 \nu
\\
\iota_2
\end{array}
\right)
\label{mono-left-sym}
\end{equation}
\begin{equation}
\vec{d}_{-\circ, j}^{\rm \;II} =\sum_{\nu=\pm} 
\left(
\xi^{(+)}_{\nu, {\rm II}}\Omega^{\iota_1\nu (j-N)}+ 
\xi^{(-)}_{\nu, {\rm II}} \Omega^{\iota_1\nu (N-j)}
\right)
\left(
\begin{array}{c}
 \nu\iota_2
\\
1
\end{array}
\right)
\label{mono-right-sym}
\end{equation}
 for clarity, where $\Omega=e^{i\frac{2}{3}\pi}$. We ommit $1/\sqrt{J}$ and replace $\tilde{\lambda}_{\nu,\circ}$ and $\tilde{f}_{\nu,\circ}$ with $\Omega$
 and $\nu$, respectively. 
In Fig.~\ref{Fig-sym},  $(\iota_1,\iota_2)_\circ$ denotes the system defined by Eqs.~(\ref{mono-left-sym}) and (\ref{mono-right-sym}) with 
 the layer index $\circ$ of region I and  a pair of signs $(\iota_1,\iota_2)$, where  $(+,+)_\downarrow$ is the original $\uparrow$ junction.
 The $x$-axis inversion transforms $(+,+)_\downarrow$  into $(-,+)_\uparrow$, replacing $j$ with $N-j$ in Eqs.~(\ref{mono-left-sym})  and
  (\ref{mono-right-sym}). 
 Here, we fix labels I and II but change ($\downarrow$,$\uparrow$) allocation such that the direction from $\downarrow$  to $\uparrow$  aligns with the positive direction of the $x$-axis. 
 If the $z$-axis is taken to point from the $\downarrow$ layer to the $\uparrow$ layer, this operation is equivalent to a $\pi$ rotation 
 about the $y$-axis.
 Comparing $(+,+)_\downarrow$ and $(-,+)_\uparrow$, $E$ is common, but $\varepsilon$ has opposite signs,
  where $\varepsilon$ denotes the site energy of the $\uparrow$ side.
  Under the chiral operation $(A_\circ, B_\circ) \rightarrow  (-\circ A_\circ, \circ B_\circ)$,
  $(-,+)_\uparrow$ changes into $(-,-)_\uparrow$,  with the reversal of $E$ and $\varepsilon$ signs.
  $(-,-)_\uparrow$ has the same $\varepsilon$
   sign as the original system $(+,+)_\downarrow$  because the sign changes two times.
  The transmission and reflection in $(+,+)_\downarrow$ is represented
  by
  \begin{equation}
\left(
\begin{array}{c}
 \vec{\xi}_{\rm I} ^{\;(-)} \\
 \vec{\xi}_{\rm II} ^{\;(+)} \\
\end{array}
\right)
=
S(E,\varepsilon)
\left(
\begin{array}{c}
 \vec{\xi}_{\rm I} ^{\;(+)} \\
 \vec{\xi}_{\rm II} ^{\:(-)} \\
\end{array}
\right),
\label{sym-system-1}
\end{equation}
where $\;^t\vec{\xi}_{\rm I}^{\;(\pm)}=(\xi_{\pm ,{\rm I} }^{(\pm)}, \xi_{\mp,{\rm I}}^{(\pm)})$, and $\;^t\vec{\xi}_{\rm II}^{\;(\pm)}=(\xi_{\pm ,{\rm II} }^{(\pm)}, \xi_{\mp,{\rm II}}^{(\pm)})$.
We define  the rows and columns of the $S$ matrix as
\begin{equation}
S=
\left(
\begin{array}{cccc}
S_{-\downarrow,+\downarrow}& S_{-\downarrow,-\downarrow} & S_{-\downarrow,-\uparrow} & S_{-\downarrow,+\uparrow}  \\
S_{+\downarrow,+\downarrow} & S_{+\downarrow,-\downarrow} & S_{+\downarrow,-\uparrow} & S_{+\downarrow,+\uparrow}  \\
 S_{+\uparrow,+\downarrow} & S_{+\uparrow,-\downarrow}& S_{+\uparrow,-\uparrow} & S_{+\uparrow,+\uparrow}   \\
 S_{-\uparrow,+\downarrow} & S_{-\uparrow,-\downarrow} &S_{-\uparrow,-\uparrow} & S_{-\uparrow,+\uparrow}  \
\end{array}
\right),
\label{sym-Smat}
\end{equation}
 where indexes $\pm \circ$ signifies the valley $K_\pm$ of the layer $\circ$.
When $\bullet=\uparrow$, Eq.~(\ref{multiple-T}) corresponds to the $\uparrow\downarrow$ block of
 Eq.~(\ref{sym-Smat}). 
 Irrespective of labels I and II, amplitude $\xi_\nu^{(+)}$  ($\xi_\nu^{(-)}$) corresponds to the $K_{\nu \iota_1}$ ($K_{-\nu \iota_1}$) valley of which the sign of the probability flow is $\nu\iota_1\iota_2$ ($-\nu\iota_1\iota_2)$.
  The arrows of wavy lines in Fig.~\ref{Fig-sym} indicate the direction of the probability flow.
 Accordingly, we obtain
\begin{equation}
\left(
\begin{array}{c}
 \vec{\xi}_{\rm II} ^{\;(+)} \\
 \vec{\xi}_{\rm I} ^{\;(-)} \\
\end{array}
\right)
=
S(E,-\varepsilon)
\left(
\begin{array}{c}
 \vec{\xi}_{\rm II} ^{\;(-)} \\
 \vec{\xi}_{\rm I} ^{\;(+)} \\
\end{array}
\right)
\label{sym-system-2}
\end{equation}
for $(-,+)_\uparrow$, and
\begin{equation}
\left(
\begin{array}{c}
 \sigma_x\vec{\xi}_{\rm II} ^{\;(-)} \\
 \sigma_x \vec{\xi}_{\rm I} ^{\;(+)} \\
\end{array}
\right)
=
S(-E,\varepsilon)
\left(
\begin{array}{c}
 \sigma_x\vec{\xi}_{\rm II} ^{\;(+)} \\
 \sigma_x\vec{\xi}_{\rm I} ^{\;(-)} \\
\end{array}
\right)
\label{sym-system-3}
\end{equation}
for $(-,-)_\uparrow$, where $\sigma_x$ stands for the Pauli matrix.
Since $S$ is unitary, Eqs.~(\ref{sym-system-1}) and (\ref{sym-system-3})
 indicate
 \begin{equation}
S(-E,\varepsilon)=
\left(
\begin{array}{cc}
 0 & \sigma_x\\
 \sigma_x& 0\\
\end{array}
\right)
\;^tS^*(E,\varepsilon)
\left(
\begin{array}{cc}
 0 & \sigma_x\\
 \sigma_x& 0\\
\end{array}
\right)
\end{equation}
leading to
\begin{equation}
T_{\mp,\pm}(-E,\varepsilon)=T_{\mp,\pm}(E,\varepsilon)
\label{UD-E-sym}
\end{equation}
 for the $\uparrow$ junction.
Equation (\ref{zeta}) possesses
  the energy symmetry
\begin{equation}
  \zeta_{\downarrow,\pm}(-E,\varepsilon)=\zeta_{\uparrow,\pm}(E,\varepsilon)^*,
  \label{zeta-E-sym}
\end{equation}
$\lambda$ is an even function of $E$, and thus Eq.~(\ref{T-smat}) for the $\uparrow$ junction  satisfies Eq.~(\ref{UD-E-sym}).

\begin{figure}
\begin{center}
\includegraphics[width=0.8\linewidth]{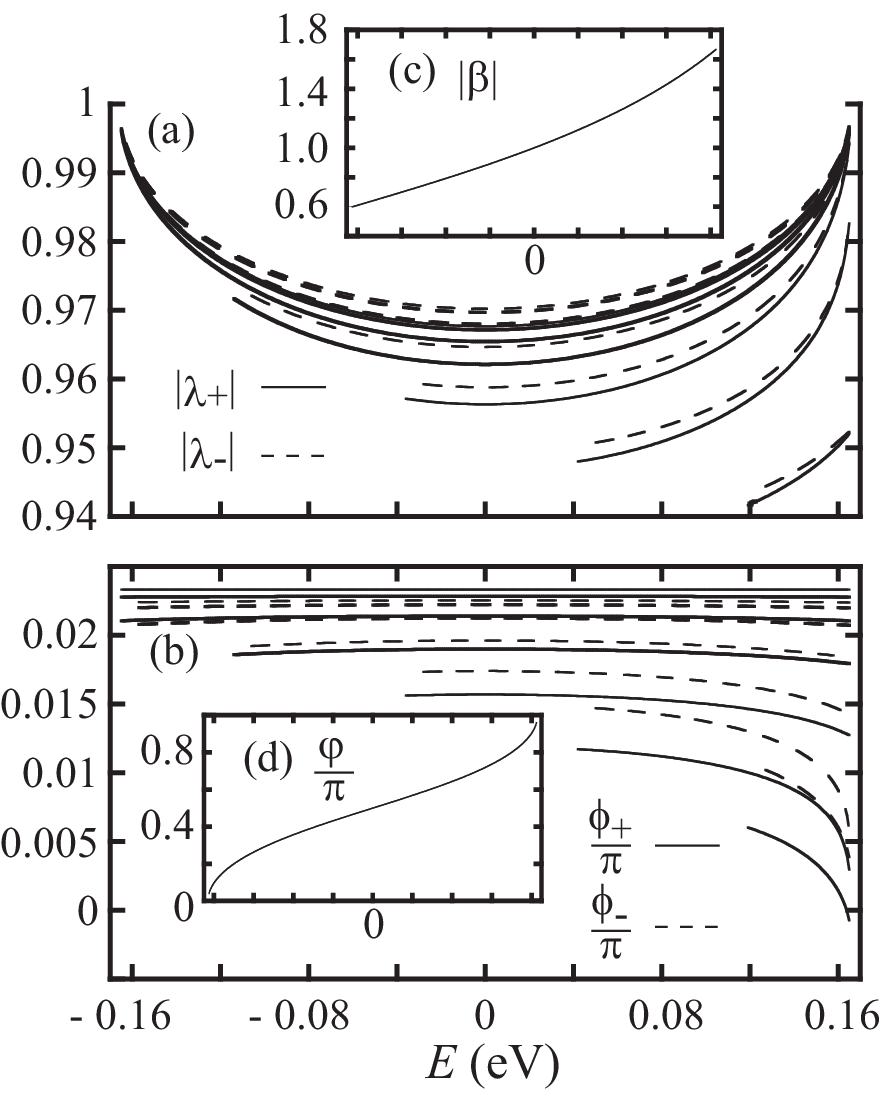}
\caption{Main Panel: (a) the absolute value $|\lambda_\pm|$  and  (b) phase $\phi_\pm$
 of the Bloch factot $\lambda_\pm=|\lambda_\pm|\exp\left(i\phi_\pm \pm i\frac{2}{3}\pi\right)$ 
 for $\varepsilon=0.35$~eV and $\sqrt{3}k_ya=0.016m\pi$,  where $m=0,1,\cdots 6$.
 In the main panel, the data are plotted over the range $s< |E+\varepsilon|$, within which Eq.~(\ref{mono-left}) describes a propagating wave.
 Both $\phi_\pm$ and  $|\lambda_\pm|$  decrease with $k_y$.
The solid (dashed) line corresponds to $\lambda_+$ $(\lambda_-)$.
 Lines  $m=0$  and $m=1$ are nearly identical in (a).
Inset:  (c) the absolute value $|\beta|$ and (d) phase $\varphi$ of the interlayer wave function ratio $\beta$.
The horizontal axis range is identical in all panels and insets.
}
\label{Fig-k}
\end{center}
\end{figure}

  \begin{figure}
\begin{center}
\includegraphics[width=0.8\linewidth]{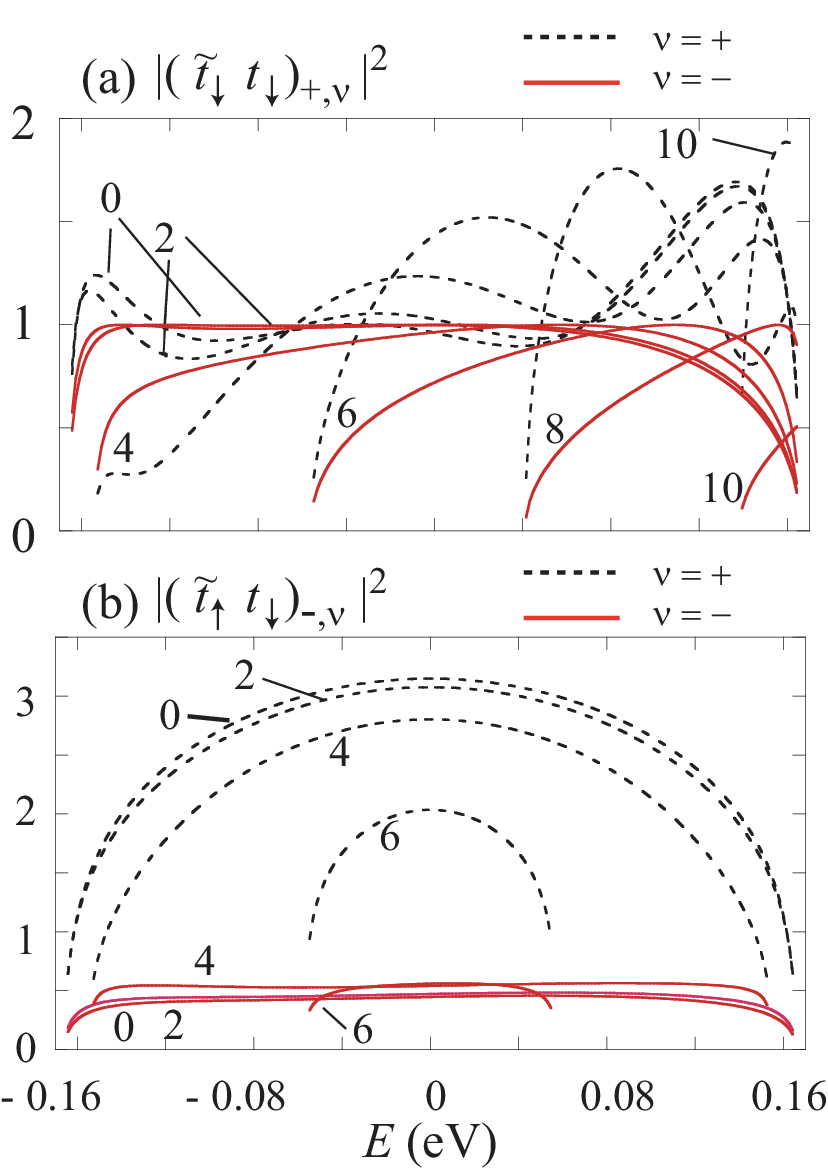}
\caption{
(a) $C^{(\downarrow)}_{\nu',\nu} \equiv \left|(\tilde{t}_\downarrow t_\downarrow)_{\nu',\nu}\right|^2$
 and (b) $C^{(\uparrow)}_{\nu',\nu} \equiv \left|(\tilde{t}_\uparrow t_\downarrow)_{\nu',\nu}\right|^2$
  are shown for $\varepsilon =0.35$~eV and $\sqrt{3}k_ya=0.02m\pi$ $(m=0,1,2,\cdots)$
in the range where $s$  is less than Eq.~(\ref{smax}).
The numbers attached to each curve represent values of $2m$.
 The $\gamma_1$-model is used.
Since $C^{(\downarrow)}_{-,+}=C^{(\downarrow)}_{+,-}$,  $C^{(\downarrow)}_{-,-}\simeq C^{(\downarrow)}_{+,+}$
  and $C^{(\uparrow)}_{+,\pm} \simeq C^{(\uparrow)}_{-,\mp}$,  we omit $C^{(\downarrow)}_{-,\nu}$ and 
 $C^{(\uparrow)}_{+,\nu}$.
}
\label{Fig-tt}
\end{center}
\end{figure}

\section{Results}
 The analytical expression becomes less accurate as $|\varepsilon|$ decreases, which is due to the condition 
$|\varepsilon| \gg \gamma_1$  under which it was obtained. 
However, when $\varepsilon$ changes continuously,  the transmission probability does not undergo discontinuous changes.
Numerical calculations are necessary to verify how large $|\varepsilon|$ must be. 
Below, we present the results for $\varepsilon = 0.35$ eV, slightly smaller than $\gamma_1$.
The corresponding gap width $2\Delta$ is about 0.33 eV.
The energy gaps reported in experiments are 0.14~eV \cite{29,30} , 0.16~eV \cite{90} , and 0.25~eV \cite{89}. In density functional theory (DFT) calculations that include structural relaxation, the energy gap increases with an electric field but saturates at 0.326~eV \cite{91}. Since DFT calculations tend to underestimate energy gaps, an energy gap of 0.33~eV is feasible.
The electric double-layer transistors technology may apply for applying a strong vertical electric field \cite{91-2}.
We employed the method in Ref. \cite{92} for the exact numerical calculations.
If the analyzed quantity is an even function of $k_y$, we implicitly restrict $k_y$  to non-negative values.

\subsection{Validity of Analytical Expressions}
Main panels of Fig.~\ref{Fig-k} show $|\lambda_\pm|$  and the phase shift $\phi_\pm$ defined in Eq.~(\ref{lambda-phase}),
 indicating that $\phi_+ \simeq \phi_- \simeq 0 $ and $\lambda_\tau\ \simeq e^{i\tau\frac{2}{3}\pi}$.
 Although  $|\lambda_+|$ is very close to $|\lambda_-|$, they differ slightly.
 Both $\phi_\pm$ and  $|\lambda_\pm|$  decrease with $k_y$.
 In the barrier region, the wave function decays very slowly ($|\lambda| \simeq 1$), and its phase closely corresponds to that of the valley states.
 When Eq.~(\ref{bi-wf}) and $k_y=0$  are substituted into Eq.~(\ref{TB-eq}), 
 the skew interlayer transfer intergrals $\gamma_3$ and $\gamma_4$ are multiplied by a factor $1+\lambda+\lambda^{-1}$.
 Since this factor is small, the analytical expressions remain valid within the $\gamma_1\gamma_3\gamma_4$-model.
 References \cite{33, 34,64} have also confirmed the effectiveness of the $\gamma_1$-model.

The factor given in  Eq.~(\ref{Re-beta}) appears both when $k_y$   is finite (Sec. III) and when 
 $k_y$ is zero (Sec. IV).
This is reminiscent of the approximation using Eq.~(\ref{def-lambda})  in Eqs.~(\ref{Z-dd}) and (\ref{Z-ud}) for finite $k_y$.
To confirm the validity of this approximation, Fig.~\ref{Fig-tt} shows the $k_y$- and $E$-dependence of $C^{(\bullet)}_{\nu',\nu} \equiv \left|(\tilde{t}_\bullet t_\downarrow)_{\nu',\nu}\right|^2$ using the $\gamma_1$-model \cite{note}.
 This factor is extracted from Eq.~(\ref{direct-T}) as the part independent of $N$. 
When $k_y$ is close to zero,  the $k_y$-dependence of the $C^{(\bullet)}$  is small.
As $s$ deviates from zero, Eq.~(\ref{direct-T}) diminishes due to the $|\lambda|$ decrease.
Since the large $T$ governs the transport, 
 Fig.~\ref{Fig-tt}  justifies
 the approximation of representing the effects of $k_y$
  solely through $\lambda$ by using Eqs.~(\ref{Z-dd}) and (\ref{Z-ud}).

\begin{figure}
\begin{center}
\includegraphics[width=0.8\linewidth]{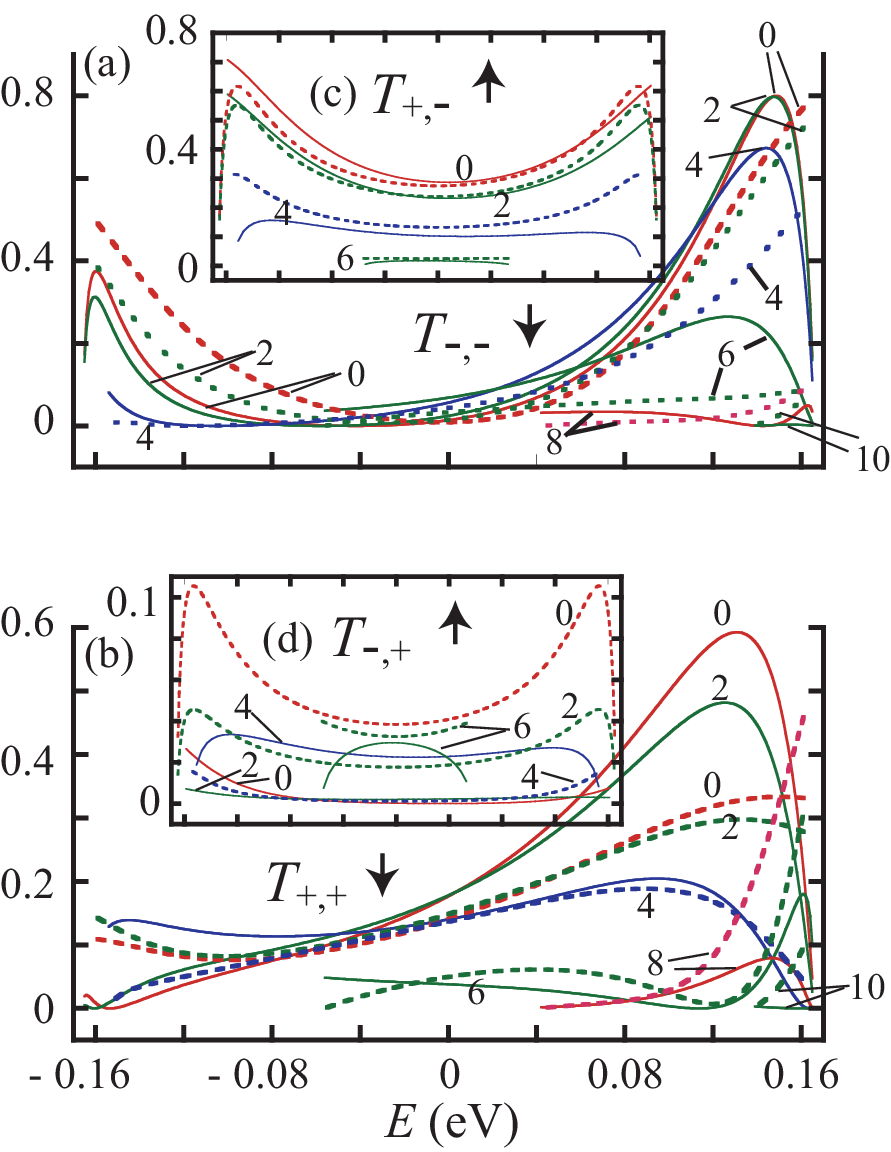}
\caption{
Main panel: (a) $T_{-,-}$, and (b) $T_{+,+}$ for the $\downarrow$ junction.
Inset: (c) $T_{+,-}$, and (d) $T_{-,+}$ for the $\uparrow$ junction. 
The attached numbers have the same meaning as in Fig.  \protect\ref{Fig-tt}.
The bilayer length $N$ is set to 56.
The solid lines represent the exact results obtained using the $\gamma_1\gamma_3\gamma_4$-model, and the dashed lines represent the analytical expressions.
The horizontal axis range is identical in all panels and insets.
}
\label{Fig-tk}
\end{center}
\end{figure}

 \begin{figure*}
%\begin{center}
\includegraphics[width=0.8\linewidth]{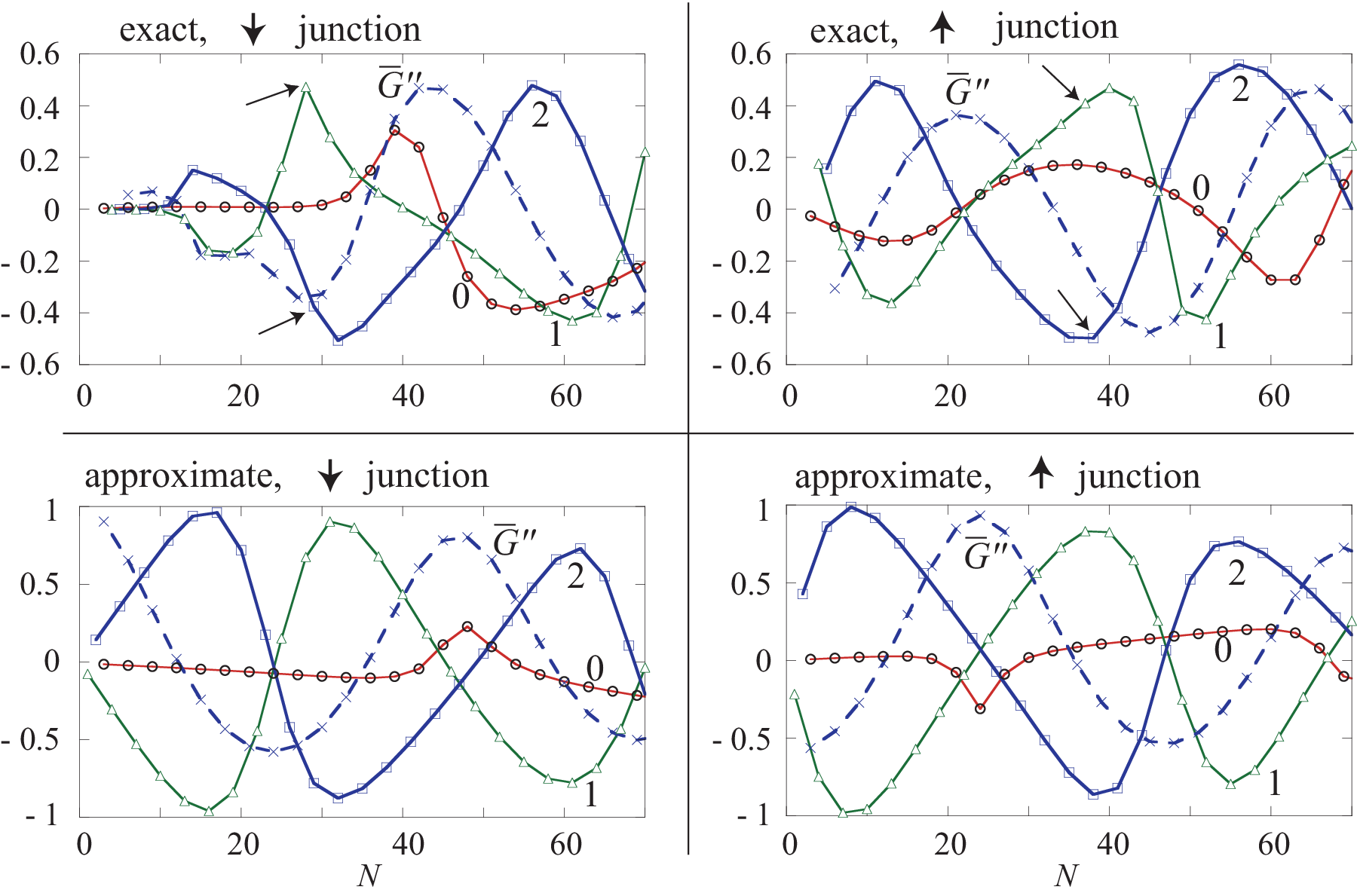}
\caption{For zero energy, the solid lines represent the normalized valley conductance $\overline{G}_v$, and the dashed lines represent the normalized curvature $\overline{G}^{\prime\prime}$ of the  $N$-$G$ curve.
The definitions of $\overline{G}_v$  and $\overline{G}^{\prime\prime}$  are provided in the main text.
Circles, triangles, and squares represent $\overline{G}_v$ for cases where 
mod$(N)=0$, mod$(N)=1$, and mod$(N)=2$, respectively.
The system width was set to $N_y=1000$.
 The left and right panels correspond to the $\downarrow$ and $\uparrow$ junctions, respectively.
 The upper and lower panels represent the exact numerical results from the $\gamma_1\gamma_3\gamma_4$-model, and the results from analytical expressions, respectively.
}
\label{Fig-G}
%\end{center}
\end{figure*}

\subsection{Transmission Probability}
Figure \ref{Fig-tk} shows the dependence of the $T_{\bullet\nu,\nu}$ on $E$ and $k_y$ for the dominant path (\ref{dominant-path}).
The dashed line depicts Eq.~(\ref{T-smat}), obtained by applying Eq.~(\ref{def-lambda}) to $\lambda$, and reproduces the exact result of the $\gamma_1\gamma_3\gamma_4$-model depicted with the solid line.
The length $N$ of the bilayer region is 56, sufficiently long for Eq.~(\ref{direct-T}) to be valid.
The most notable difference between the two types of junctions is that energy symmetry is present only in the $\uparrow$ junction.
This is attributable to $\lambda^N$ appearing alone in Eq.~(\ref{Z-ud}), whereas in Eq.~(\ref{Z-dd}), it appears alongside $\beta^{-1}$.
As shown in Fig.~\ref{Fig-k}, the symmetry of pristine bilayer graphene is reflected in the fact that $\lambda$ is an even function of energy $E$, whereas $\beta$ is not.
As indicated by Eqs.~(\ref{beta}) and (\ref{def-lambda}), exchanging the $\uparrow$ and $\downarrow$ layers -- i.e., reversing the sign of  $\varepsilon$ -- does not affect $\lambda$, but inverts $\beta$ as $\beta(-\varepsilon)=1/\beta(\varepsilon)$.
In contrast,  the chiral operation simultaneously inverts the signs of both  $E$ and $\varepsilon$,
 leaving $\lambda$ unchanged but flipping the sign of $\beta$.
By combining these pristine bilayer graphene properties with the POG symmetry discussed in Sec. V, we can explain why the $N$-dependent factor in the $\uparrow$ junction, as expressed in Eq.~(\ref{Z-ud}), is free from the influence of $\beta$. We refer to this phenomenon as self-cancellation of $\beta$.

When $N$ becomes too large, the transmission probability within the band gap approaches zero, regardless of the type of junction.
Here, we focus on the case  $N<100$.
Fig.~\ref{Fig-k}  (b) shows that the phase $\phi$ of the Bloch factor $\lambda$ varies by less than 
$0.001\pi$  with energy $E$ in the range $|E| < {\rm min}(\varepsilon-s|\gamma_0|, \Delta)$, which is relevant for the $\uparrow$ junction, 
 as given by Eq.~(\ref{smax}).
Therefore, when $N<100$, the variation of  $\theta_{\tau,\uparrow}$ with energy is less than $0.1\pi$. This implies that if  $\cos^2\theta_{\tau,\uparrow}$ is small at 
$E=0$, i.e., when $\theta_{\tau,\uparrow}$  is close to an odd multiple of  $\pi/2$, Eq.~(\ref{T-smat}) is suppressed across the entire gap $|E| < \Delta$ for the $\uparrow$ junction.
Figure \ref{Fig-tk} (d) shows an example of such suppression.
The agreement between the solid and dashed lines in Fig.~\ref{Fig-tk} (d) appears to be worse compared to Fig.~\ref{Fig-tk} (c).
However, the critical point is that the $T_{-,+}$ remains below 0.1 across the entire gap.
This is closely related to the energy range where a VB occurs.
For the $\uparrow$ junction, Eq.~(\ref{Re-beta}) is independent of $\varphi$, and $\varphi$ does not shield the energy dependence of 
 $|\lambda|$.  Both $T_{+,-}$ in Fig.~\ref{Fig-tk} (c) and $|\lambda_-|$ in Fig.~\ref{Fig-k} (a) decrease as $E$ approaches zero.
In contrast, as shown in Fig.~\ref{Fig-k} (d),  $\varphi$ and $\theta_{\tau,\downarrow}=
\theta_{\tau,\uparrow}-\varphi$  vary by $\pi$ across the gap, resulting in zeros of 
$\cos\theta_{\tau,\downarrow}$ and hence of Eq.~(\ref{T-smat}) for the $\downarrow$ junction within the gap region.
These zeros appear around  $E=-0.05$~eV in Fig.~\ref{Fig-tk}  (a) and near the gap edges 
 in Fig.~\ref{Fig-tk} (b), suppressing  the $T_{\pm,\pm}$.  Due to the significant change in $\varphi$  within the gap, the suppression of the $T$ across the entire gap cannot occur in the $\downarrow$ junction.

\subsection{$N$-Dependence of Conductance at Zero Energy}
The valley-resolved conductance $G_{\nu',\nu}$ is a critical parameter for understanding electronic transport properties. Using Landauer's formula \cite{93}, it is calculated by summing the transmission probabilities over $k_y$: $G_{\nu',\nu}=\frac{2e^2}{h}\sum_{m=-M}^M T_{\nu',\nu}(m\Delta k_y)$. 
The periodic boundary condition in the $y$-direction leads to the discretization of $\Delta k_y=\frac{2\pi}{\sqrt{3}a N_y}$ with the system width $\sqrt{3}aN_y$.
In our computations, $N_y$ is taken as 1000.
Equation (\ref{smax}) determines $M$, the maximum allowable integer $m$, as $s_{\rm max}=\sin(M\Delta k_y)$.
The channel number per valley is then $2M+1$. Figure \ref{Fig-G} shows 
 the normalized valley conductance $\overline{G}_v=\frac{G_v}{G}$ at the gap center
 $E=0$, where $G=\sum_{\nu',\nu}G_{\nu',\nu}$ and $G_v=\sum_{\nu}(G_{+,\nu}-G_{-,\nu})$ represent the charge conductance and valley conducance, respectively. 
  As $|\phi| \ll 1$,  Eq.~(\ref{Re-beta}) induces oscillations in Eq.~(\ref{T-smat}) with a period of three in $N$. 
  Here, we define mod$(N)$ as the remainder of $N$ mod 3. The data $\overline{G}_v$  in Fig.~\ref{Fig-G}  are grouped by mod$(N)$, and connected by lines within each group.
  The left and right panels correspond to the $\downarrow$ and $\uparrow$ junctions, respectively.
  The lower (upper) panels present the analytic expressions (the exact numerical results from the $\gamma_1\gamma_3\gamma_4$-model). 
  The minor transmission channels -- $T_{\mp \bullet,\pm}$ in the $\bullet$ junction  -- are neglected in the lower panel. However, they are included in the upper panel, leading to a reduction in the amplitude.
   Additionally,  for small  $N$, terms with $n \geq 1$ in Eq.~(\ref{multiple-T})  become significant and cannot be ignored.
   This is why the upper and lower panels discrepancy is particularly pronounced for $N<20$ in the $\downarrow$ junction.
  Except for this point, the analytical expressions closely replicate the oscillation period and the positions of the maxima and minima in the exact results when mod$(N) \neq 0$.
 Equation (\ref{Re-beta-theta}) changes by $\Delta \theta = \left(\frac{4}{3}\frac{\varepsilon}{|\varepsilon|}\pi+\phi_{\frac{\varepsilon}{|\varepsilon|}}-\phi_{-\frac{\varepsilon}{|\varepsilon|}}\right)N$ when the incident valley changes from $K_-$ to $K_+$.
 This  $\Delta \theta$ introduces a significant difference between $T_{\bullet,+}$
  and $T_{-\bullet,-}$ in the $\bullet$ junction, except when $\Delta \theta$  is close to an integer multiple of $\pi$.
  This explains the differences between $T_{-,-}$ and $T_{+,+}$ and why $T_{+,-}$  is much larger than $T_{-,+}$  in Fig.~\ref{Fig-tk}.
Reversing the sign of $\varepsilon$ changes $\Delta\theta$ and hence the $G_v$. 
The VF in zigzag graphene ribbons also reverses its polarity depending on the sign of the site energy \cite{52, 59}.
When $ |\phi_+-\phi_-|N$ is small, $|\Delta \theta|$  can be approximated as $\frac{4}{3}\pi N$.
 This accounts for the smaller amplitude of the solid line for mod$(N)=0$ compared to mod$(N)\neq 0$ in Fig.~\ref{Fig-G}.

Although Eq.~(\ref{dominant-path})  indicates that the valley reversal occurs only in the $\uparrow$ junction, both the junctions exhibit $G_v$ with the same sign in Fig.~\ref{Fig-G}.
At zero energy, $\beta$ equals $i$, causing a phase shift of $\varphi=\frac{\pi}{2}$  in Eq.~(\ref{Z-dd}) relative to Eq.~(\ref{Z-ud}). 
Since the transmission probability is proportional to the square of Eqs.~(\ref{Z-dd}) and (\ref{Z-ud}), the sign reversal occurs, 
canceling out the effect of the valley reversal.
However, as $E$ moves away from zero,  $\beta$ deviates from $i$, causing this cancellation to become imperfect.
At zero energy, differences due to the type of junction do not manifest in valley selectivity in this way, but they do appear in the curvature of the $N$-$G$ curve,   $G^{\prime\prime}(N)=G(N+1)+G(N-1)-2G(N)$.
The dashed line in Fig.~\ref{Fig-G} represents the normalized curvature  $\overline{G}^{\prime\prime}(N)=G''(N)/[G(N+1)+G(N-1)+2G(N)]$ for the case where mod($N$) =0.
By applying $\varphi=\frac{\pi}{2}$ and  $\phi_+\simeq \phi_-$ to Eq.~(\ref{Re-beta}), 
the resulting $\overline{G}''=\bullet 3\cos(2N\phi)/(4-\bullet\cos(2N\phi))$ acquires opposite signs in the $\downarrow$ and $\uparrow$ junctions,
 as Fig.~\ref{Fig-G} depicts. 
 Interestingly, the phase of the $N$-$\overline{G}_v$ curve is shifted by $\frac{\pi}{2}$ compared to the $N$-$\overline{G}''$ curve, while their amplitudes are roughly equivalent.
We can predict the $N$-$\overline{G}_v$ curve by shifting the $N$-$\overline{G}''$ curve by a quarter wavelength. 
 As $|s|$ approaches $s_{\rm max}$, the difference between $\tilde{f}_{+,\circ}$ and $\tilde{f}_{-,\circ}$ diminishes, making Eq.~(\ref{dominant-path}) no longer a good connection representation.
 However,  as $|s|$ increases, $|\lambda|$ diminishes, resulting in a lower transmission probability, as illustrated in Fig.~\ref{Fig-tk}.
 Refer to Appendix A about this relation between $s$ and $|\lambda|$.
Consequently, the effect of large $|s|$ has a negligible impact on $G$,
 and the exact results show good agreement in Fig.~\ref{Fig-G}, justifying the approximation $\rho_{\tau,l} \simeq \omega^2$ used in Eq.~(\ref{xi-down-up}).

\begin{figure}
\begin{center}
\includegraphics[width=0.9\linewidth]{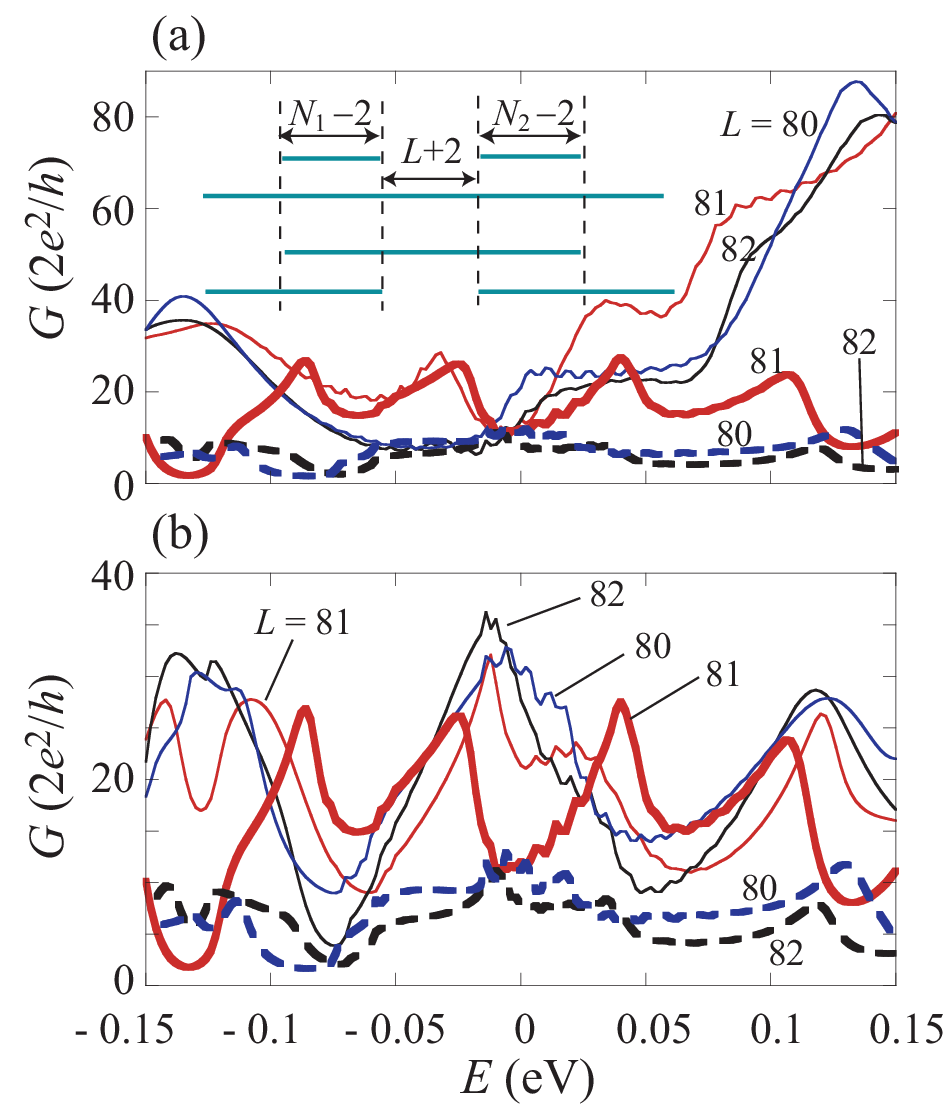}
\caption{
The conductance of double junctions is calculated using the Landauer formula for cases where the integer $L$, representing the length of the central monolayer region, is 80, 81, and 82. The attached numbers indicate $L$. As in Fig. \protect\ref{Fig-G}, $N_y=1000$.
Inset: Side view of the double junctions.
The $\uparrow$ layer is removed for  $j \leq 0$ and $N_1+N_2+L\leq j$,  where $x=\frac{a}{2}j$.
In the double $\downarrow$ ($\uparrow$)  junction, the $\uparrow$ ($\downarrow$) layer is removed in the range $N_1 \leq j \leq N_1+L$.
(a)
The thin lines correspond the double $\downarrow$ junction  ($N_1=28,N_2=29)$ and the thick lines correspond to  the double $\uparrow$ junction 
($N_1=37,N_2=38).$ 
(b)
The thick lines represent the same data as the thick lines in (a).
The thin lines correspond to the double $\uparrow$  junction $(N_1=37, N_2=39$).
}
\label{Fig-VB}
\end{center}
\end{figure}

\subsection{Valley Blockade in Double Junctions}
The charge conductance $G$ is more experimentally accessible than the valley conductance $G_v$.
As an example demonstrating the self-cancellation of $\beta$, we calculate $G$ exactly using 
the $\gamma_1\gamma_3\gamma_4$ model for a double junction with reversed-polarity VFs connected in series.
 Figure  \ref{Fig-VB} displays the $G$ in case $L=80,81,82$, where the central monolayer region has
  the length $(L+2)\frac{a}{2}$.  The two bilayer regions sandwitch the central monolayer and have the lengths  $(N_1-2)\frac{a}{2}$, and 
 $(N_2-2)\frac{a}{2}$.
 In Fig.  \ref{Fig-VB} (a), the thin lines correspond to the double $\downarrow$ junction with $(N_1, N_2)=(28,29)$, while the thick lines correspond to the double $\uparrow$ junction with $(N_1, N_2)=(37,38)$.
The corresponding $\overline{G}_v$ pointed by arrows in Fig.~\ref{Fig-G} suggest
 that the VB occurs in Fig.  \ref{Fig-VB} (a).
The VB actually occurs across the entire gap for the dashed thick lines with non-zero mod($L$).
The absence of $\varphi$ in  Eq.~(\ref{Re-beta-theta}) enables the suppression of  $T$ for only one valley across the whole gap, as shown in the insets of  Fig.  \ref{Fig-tk}.  
However, when mod($L$) =0, valley-preserving transmission, which was ignored in the analytical expression, resonates in the central monolayer, disrupting the VB in the solid thick line.
The main panels of Fig.~\ref{Fig-tk}  illustrate that the $\varphi$ limits the suppression of $T$ to an energy region much narrower than the gap width $\Delta$ in the $\downarrow$ junction.
This is why the VB is unclear in the thin lines compared to the thick dashed lines.
 In Fig.~\ref{Fig-VB} (b), thin lines represent the $G$ of the double $\uparrow$ junction with $(N_1,N_2)=(37,39)$, and 
thick lines  from Fig.~\ref{Fig-VB} (a) are re-displayed.
The two double $\uparrow$ junctions differ by only one in $N_2$.
In the case where $N_2=39$, the right $\uparrow$ junction lacks valley selectivity and thus does not exhibit VB. 
 Merely changing the thickness of the energy barrier by $\frac{a}{2}$ results in a difference in the VB. 
Regarding the effect of $L$, only whether mod($L$) is zero or not is related to the VB in double $\uparrow$ junctions. The VB does not appear in double $\downarrow$ junctions irrespective of $L$.

\subsection{Interlayer Displacement}
\begin{figure}
\begin{center}
\includegraphics[width=0.9\linewidth]{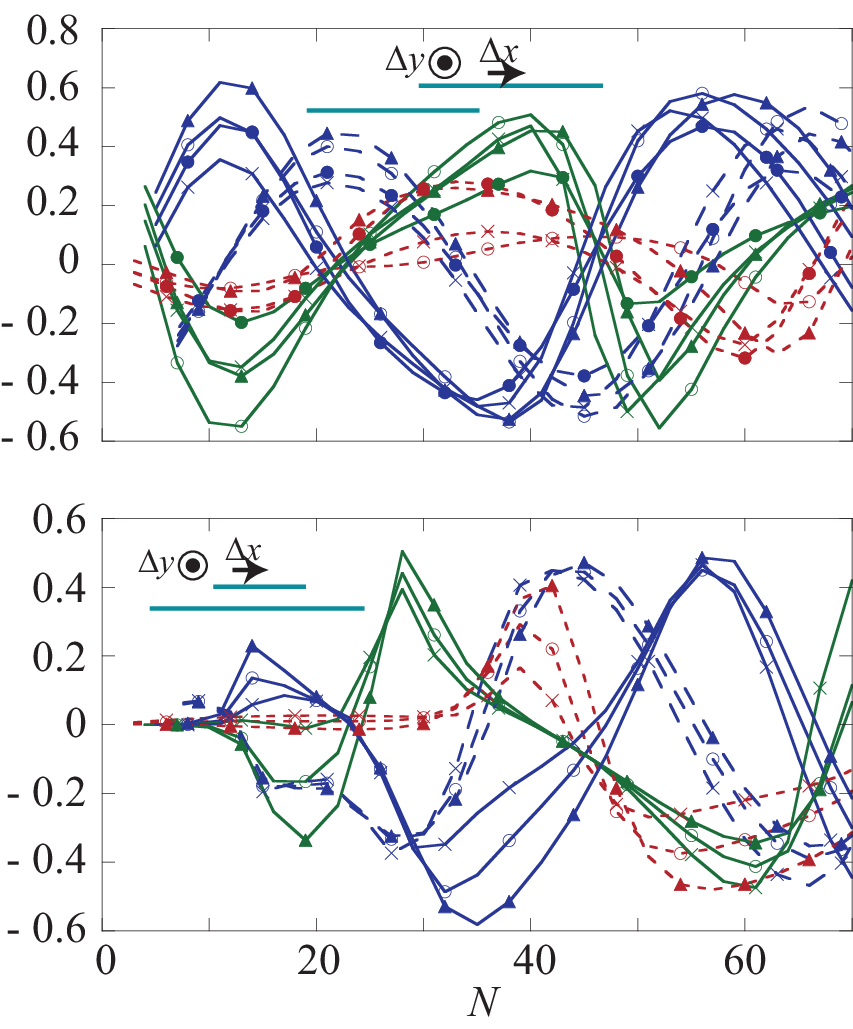}
\caption{Data corresponding to the upper panels of Fig.~\ref{Fig-G} are plotted. Open circles, solid circles, crosses, and triangles represent displacements $(b,0)$, $(-b,0)$, $(0,b)$, and $(0,-b)$, respectively, for the four displacements $(\Delta x,\Delta y) = (\pm b, 0)$ and $(0, \pm b)$, where $b = 0.01$~nm. Data points for $\overline{G}_v$ are connected by solid lines for $\mathrm{mod}(N) = 1, 2$ and by dotted lines for $\mathrm{mod}(N,3) = 0$. Since deviations from Fig.~\ref{Fig-G} are small, it is possible to distinguish between $\mathrm{mod}(N) = 1$ and $2$. Points for $\overline{G}^{\prime\prime}$ are connected by dashed lines. To avoid cluttering the figure, symbols (circles, triangles, squares) are shown only for selected data points; however, the lines connect all data points.}
\label{Fig-G-Tuika}
\end{center}
\end{figure}

\begin{figure}
\begin{center}
\includegraphics[width=0.9\linewidth]{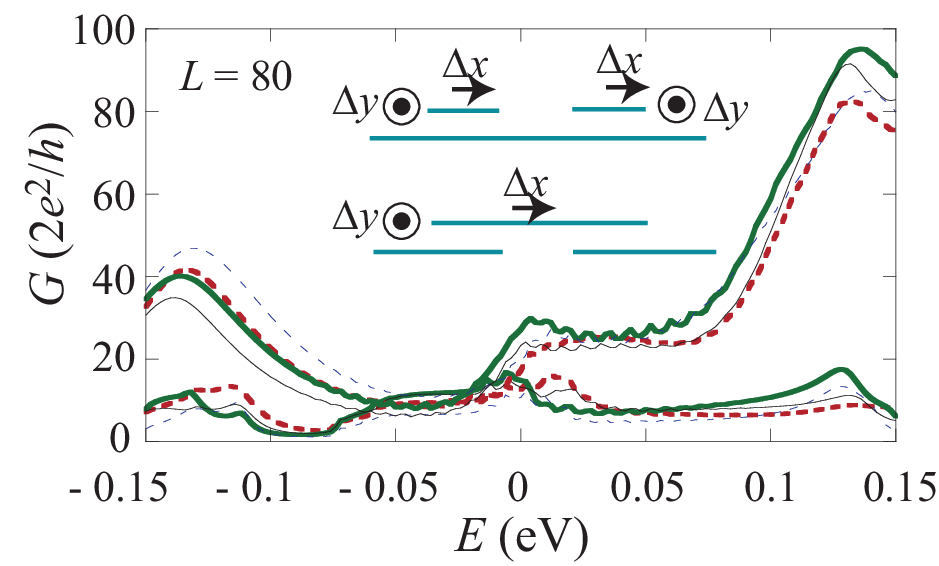}
\caption{
 Data corresponding to the case $L$=80 in Fig.~\ref{Fig-VB} (a) are plotted for the same displacements as in Fig.~\ref{Fig-G-Tuika}. In the double $\uparrow$ junction, the two  $\uparrow$  layers on the left and right sides are assumed to share the same displacement. Thick dashed and solid lines correspond to $(b,0)$ and $(-b,0)$, respectively, while thin dashed and solid lines correspond to $(0,b)$ and $(0,-b)$, respectively.
The small deviations from Fig.~\ref{Fig-VB} (a) confirm that the distinction between double $\downarrow$ and double $\uparrow$ junctions remains valid even in the presence of such displacements.
}
\label{Fig-VB-Tuika}
\end{center}
\end{figure}

The effect of a deviation from AB stacking is analyzed by expressing the displacement of the $\uparrow$ layer relative to the $\downarrow$ layer as $(\Delta x,\Delta y)$. 
We calculate the transmission probability using the interlayer transfer integrals multiplied by the factors $\exp(-dr/r_d)$, where $r_d=0.045$~nm,  and $dr$ denotes the change in interatomic distance induced by the displacement \cite{Lambin}.
We determine  $dr$ using $a=$0.246~nm and the interlayer spacing 0.335~nm.
The interlayer transfer integrals change by at most approximately 10 \% due to $(\Delta x,\Delta y)$.
Figures \ref{Fig-G-Tuika} and \ref{Fig-VB-Tuika} correspond to the upper panels of Fig.~\ref{Fig-G} and the case $L$=80 in Fig.~\ref{Fig-VB} (a), respectively, showing the results for displacements $(\Delta x,\Delta y)=(\pm b,0),$ and $(0,\pm b)$, where $b$=0.01~nm.
For the double $\downarrow$ junction, we assume that both $\uparrow$ layers on the left and right sides undergo the same displacement.
In Fig.~\ref{Fig-G-Tuika}, the displacements $(b,0)$ and  $(-b,0)$ yield identical results for the $\downarrow$ junction, but different results for the $\uparrow$  junction. 
On the other hand, for the double $\downarrow$  junction, since 
 $N_1$ and $N_2$ differ, the two displacements result in distinct outcomes.
The small effects of $(\Delta x,\Delta y)$ indicate that the analytic formulas remain valid for displacements up to the order of 0.01~nm. 
However, as the displacement increases beyond 0.01~nm, the agreement between the analytic expressions and the exact results obtained from the $\gamma_1\gamma_3\gamma_4$ model deteriorates, due to the effects of the skew transfer integrals,  $\gamma_3$ and $\gamma_4$.

\section{Discussion}
The incident electrons split into $\uparrow$ and $\downarrow$ wave components at the entrance, as shown in Eq.~(\ref{xi-j}). Simultaneously, 
 even when the electrons are incident from a single valley, they are transformed into a superposition of opposite valley states, as indicated by Eq.~(\ref{eta3-eta1}). 
When the two-layer wave components merge back into a single-layer wave and return to a single valley state at the exit, a phase shift of $\frac{\pi}{2}$  occurs compared to the entrance. 
The $\beta^{-1}$ factor appears only in the $\downarrow$ layer in Eq.~(\ref{xi-j}).
These properties are the same for both types of junctions. 
 In the $\downarrow$ junction, electrons departing from the source electrode can reach the drain electrode without traversing from one layer to the other. 
Nevertheless, they inevitably experience the effect of $\beta$ due to the aforementioned wave splitting and merging, which results in an energy asymmetry in the transmission probability.
In contrast, in the $\uparrow$ junction, interlayer electron transport is required, leading to additional effects of  $\beta$ compared to the $\downarrow$ junction.
This additional effect cancels out the $\beta$ effect  present in the $\downarrow$ junction.
This self-cancellation of $\beta$ is enforced by the energy symmetry inherent in the $\uparrow$ junction.

The transmission probability given in Eq.~(\ref{T-smat}) is expressed as the product of an 
$N$-dependent factor and an $N$-independent factor.
The energy  symmetry must hold for each factor individually.
 The  factors $\lambda$ and $\beta$ determine the $N$-dependent part, and thus the energy dependece of their phases $\phi$ and $\varphi$ is cruicial.
 We consider the case $N<100$, where the phase $N\phi$ varies only slightly with $E$. Thus, if transmission is suppressed at $E=0$, it remains low across the gap in the $\uparrow$ junction.
In contrast, the phase $\varphi$ of $\beta$ varies by $\pi$ 
within the gap (Fig.~\ref{Fig-k} (d)), making such suppression impossible in the $\downarrow$ junction. 
When $|\lambda|=1$, $\phi$ corresponds to the wave vector component $k_x$, 
and the situation where $E$ depends only weakly on $k_x$ is referred to as a nearly flat band. 
In contrast, Fig.~\ref{Fig-k} (b) exhibits a nearly anti-flat band, where $\phi$ depends only weakly on $E$.
The emergence of a full-gap VB in the double $\uparrow$ junction (Fig.~\ref{Fig-VB} (a)) serves as direct evidence of  this
 anti-flat band and the self-cancellation of $\beta$.

The system's width $N_y$ is 1000 in  Figs.~\ref{Fig-G} and \ref{Fig-VB}. 
Since $k_y$  is discretized with a spacing $\Delta k_y=2\pi/(\sqrt{3}aN_y)$, 
the channel number changes discontinuously each time $E$ changes by $|\gamma_0|\frac{\sqrt{3}}{2}a\Delta k_y$.
 This leads to the fine oscillations with a period of about 0.01~eV in Fig.~\ref{Fig-VB}.
 The amplitude of these fine oscillations is, at most approximately $2e^2/h$,  demonstrating that the chosen $N_y$  is sufficiently large for $\frac{G}{N_y}$ to become independent of $N_y$.
In Ref. \cite{94}, side electrodes were proposed for decomposing $G$ into $k_y$ components to realize VF. For the VB discussed here, $k_y$ decomposition is unnecessary.

In the $pn$ junction of a zigzag graphene ribbon, valley-reversed transmission similar to that in the $\uparrow$ junction also occurs, but it is 
 irrelevant to $\beta$ due to the monolayer structure \cite{52,59}.
We assume a defect-free armchair edge at the boundary between the monolayer and bilayer regions.
Defects introduced at this boundary degrade the VF.
For example, if a carbon dimer is adsorbed at the boundary, the local increase in $N$ renders $N$ undefined.
Reference \cite{64} demonstrated that as long as the line density of these carbon dimers is low and 
$N$ remains nearly constant, the VF survives.
Bottom-up synthesis methods based on chemical synthesis, which are effective in producing edges with minimal defects, are rapidly advancing and are expected to enable the detection of $\phi$ and $\varphi$ proposed in this paper \cite{95, 96}.

\section{Conclusion}
We derived analytical expressions for the transmission probabilities of two types ($\downarrow$ and $\uparrow$) of partially overlapped graphene (POG) with AB stacking in the energy gap region $|E| < \Delta$, where a vertical electric field induces the gap. The derivation is based 
 on the $\gamma_1$-model, under the assumption that the half-gap width $\Delta$ is comparable to the vertical interlayer transfer integral $\gamma_1$.
As the transverse wave vector $k_y$ increases from zero, the decay length in the longitudinal (transport) direction decreases, making transmission channels with small $k_y$ the dominant contributors. The analytical expressions were validated by comparison with numerical results obtained from the $\gamma_1\gamma_3\gamma_4$-model. We found that valley-conserving (valley-reversing) transmission dominates in the $\downarrow$ ($\uparrow$) junction.
We further computed the $N$-$G_v$ and $N$-$G$ curves at the center of the energy gap, where $G_v$ and $G$ represent the valley and charge conductances, respectively, and $N$ denotes the bilayer length as $(N-2)$ units of the half lattice constant $\frac{a}{2}$. While the sign of $G_v$ remains largely unchanged between the two junction types, the curvature of the $N$-$G$ curve exhibits opposite signs.
Although the analytical expressions remain valid even when the interlayer displacement is about 0.01~nm, the effects of the skew interlayer transfer integrals can become more relevant compared to the case of perfect AB stacking.

The POG structure reveals the distinct energy dependence of two bilayer phases within the gap: $\phi$, the phase of the Bloch factor $\lambda$, and $\varphi$, the phase of the interlayer amplitude ratio $\beta$. While $\phi$ remains nearly constant and is an even function of energy, $\varphi$ varies significantly and lacks even symmetry.
Using symmetry arguments involving the chiral operation, $\pi$ rotation, and probability conservation, we demonstrated that the valley-reversing transmission probability in the $\uparrow$ junction is an even function of energy. This leads to two contrasting behaviors: the conductance of the $\uparrow$ junction is governed solely by $\phi$ and is therefore nearly energy-independent, whereas that of the $\downarrow$ junction depends on both $\phi$ and $\varphi$, resulting in strong energy dependence.
While $\phi$ and $\varphi$ stem from the intrinsic electronic structure of pristine bilayer graphene, their distinct energy dependence remains inaccessible in its pristine form. It is only through the introduction of the POG configuration that these phase effects become observable 
 through conductance. This difference between the $\downarrow$ and $\uparrow$ junctions is interpreted as a self-cancellation of the $\beta$ effect: the additional $\beta$ contribution appearing in the $\uparrow$ junction cancels the original contribution present in the $\downarrow$ junction. The occurrence of valley blockade across the entire gap exclusively in the double $\uparrow$ junction provides strong evidence for this self-cancellation mechanism.
These findings highlight the role of POG structures in revealing otherwise hidden features of bilayer graphene and provide a clear theoretical framework for interpreting the energy dependence of valley and charge transport within the band gap.

%\section{\label{sec:level1}First-level heading:\protect\\ The linebreak was forced \lowercase{via} \textbackslash\textbackslash}
\appendix
\section{Note on Eq.~(12)}
In the square root  $\sqrt{A}=e^{i\frac{\theta}{2}}\sqrt{|A|}$ of a complex number $A=|A|e^{i\theta}$, we define the range of the phase $\theta$
 as $-\pi < \theta \leq \pi$. Under this standard definition, 
\begin{equation}
\sqrt{-1+o_1\pm io_2} \simeq \frac{o_2}{2}\pm i\left(1 -\frac{o_1}{2}\right)
\label{app-root}
\end{equation}
 holds when $|o_1| \ll 1$ and $0< o_2 \ll 1$.
 Using Eqs.~(\ref{def-lambda}) and (\ref{app-root})  under the condition that $|\varepsilon|, |E| , \gamma_1 \ll 1 (=|\gamma_0|)$, 
 we obtain
 \begin{equation}
\lambda_{\tau,l} \simeq \left(1- \frac{ \chi_--li\chi_+}{\sqrt{3+s^2}}\right)\left(-\frac{c}{2}+\tau l i \frac{\sqrt{3+s^2}}{2}\right)
\end{equation}
where $c=\sqrt{1-s^2}$,
\begin{equation}
\sqrt{p-s^2+iq} =\chi_++i\chi_-,
\end{equation}
 and 
\begin{equation}
\chi_\pm=\frac{1}{\sqrt{2}}\sqrt{\sqrt{(p-s^2)^2+q^2}\pm p \mp s^2 }.
\end{equation}
Since $|s|$ cannot exceed Eq.~(\ref{smax}), 
$|\lambda_{\tau,l}|\simeq |\lambda_\tau| e^{il\tau \frac{2}{3}\pi}$, and 
 $|\lambda_{\tau}|\simeq 1-\frac{2\chi_-}{\sqrt{3+s^2}}< 1$.
An increase in $|s|$ leads to an increase in $\chi_-$, which in turn causes a decrease in $|\lambda_{\tau,l}|$.

\section{Scattring matrixes for the $\downarrow$ and $\uparrow$ junctions}
Using matrixes $S_\downarrow$ and $S_\uparrow$ of Refs. \cite{34} and \cite{64}, we derive Eq.~(\ref{T-smat}).
$S_\downarrow$ is responsible to the left transition $(j=0)$ as
\begin{equation}
\left(
\begin{array}{c}
\vec{\eta}^{\;(+)}
\\
\vec{\xi}^{\;(-)}_{\rm I}
\end{array}
\right)
=S_\downarrow
%\left(
%\begin{array}{cc}
%r_\downarrow
%& 
%t_\downarrow\\
%\tilde{t}_\downarrow
%&
%\tilde{r}_\downarrow
%\end{array}
%\right)
\left(
\begin{array}{c}
\vec{\eta}^{\;(-)}
\\
\vec{\xi}^{\;(+)}_{\rm I}
\end{array}
\right)
\end{equation}
where $\vec{\eta}^{\;(\pm)} =(\eta_{+,+}^{(\pm)},\eta_{-,+}^{(\pm)},\eta_{-,-}^{(\pm)},\eta_{+,-}^{(\pm)})$.
At the right transition $(j=N)$, 
\begin{equation}
\left(
\begin{array}{c}
\vec{\eta}^{\;(-)}
\\
\vec{\xi}^{\;(+)}_{\rm II}
\end{array}
\right)
=S_\bullet
\left(
\begin{array}{c}
\vec{\eta}^{\;(+)}
\\
\vec{\xi}^{\;(-)}_{\rm II}
\end{array}
\right).
\end{equation}
In the $\uparrow$ junction, $\bullet = \uparrow$ and $S_\bullet \neq S_\downarrow$.
In the case of zero $k_y$, Refs. \cite{34} and \cite{64} indicate that
\begin{eqnarray}
S_\downarrow
&=&
\left(
\begin{array}{ccc}
\frac{\alpha_+^2}{v_+} &
\frac{-\alpha_+\alpha_-}{v_+} &
\frac{\alpha_+}{v_+}
\\
\frac{-\alpha_+\alpha_-}{v_-} &
\frac{\alpha_-^2}{v_-} &
\frac{-\alpha_-}{v_-}
\\
\alpha_+&
-\alpha_-&
1
\\
\end{array}
\right)\otimes \frac{u_+}{\zeta_{\downarrow,+}} 
\nonumber \\
& &
+\left(
\begin{array}{ccc}
\frac{1}{v_+} &
\frac{1}{v_+} &
\frac{1}{v_+}
\\
\frac{1}{v_-} &
\frac{1}{v_-} &
\frac{1}{v_-}
\\
1&
1&
1
\\
\end{array}
\right)\otimes \frac{u_-}{\zeta_{\downarrow,-}}
-{\bf 1}_6
\label{app-S-d}
\end{eqnarray}
where ${\bf 1}_n$ denotes the $n$-dimensional identity matrix,
\begin{equation}
 u_\pm=
 \left(
\begin{array}{cc}
 1& \pm 1 \\
 \pm 1 & 1
\end{array}
\right)
\end{equation}
\begin{equation}
v_\pm =\frac{\alpha_\pm}{\beta_\mp}(\beta_--\beta_+)
\label{app-v}
\end{equation}

\begin{equation}
\zeta_{\downarrow,\pm} =1+\frac{\alpha_+^{\pm 1}\beta_-+\alpha_-^{\pm 1}\beta_+}{\beta_--\beta_+},
\label{app-zeta}
\end{equation}

\begin{equation}
\alpha_\pm =\left\{
\begin{array}{ccc}
\frac{E+\varepsilon}{\sqrt{p\pm Q}}\frac{E}{|E|} & \cdots&  \mbox{  $|\varepsilon| > |E|> \Delta$} \\
\frac{E+\varepsilon}{\sqrt{p\pm iq}}  &  \cdots & \mbox{  $\Delta> |E| $} 
\label{app-alpha}
\end{array} 
\right.
\end{equation}
\begin{equation}
\beta_\pm =\left\{
\begin{array}{ccc}
\frac{2\varepsilon E\mp Q}{\gamma_1(E-\varepsilon)} & \cdots&  \mbox{  $|\varepsilon| > |E|> \Delta$} \\
\frac{2\varepsilon E\mp iq}{\gamma_1(E-\varepsilon)}  &  \cdots & \mbox{  $\Delta> |E| $} 
\end{array} 
\right.
\label{app-beta}
\end{equation}
and  $Q=\sqrt{(4\varepsilon^2+\gamma_1^2)(E^2-\Delta^2)}$.
When $s=0$ and $|E| < \Delta$,   $f_{+,\downarrow,\pm}$ and Eq.~(\ref{beta}) coinside with Eqs.~(\ref{app-alpha}) and (\ref{app-beta}), respectively.
Replacement of $iq$ to $Q$ 
is the analytic continuation from the gap region $|E| < \Delta$
 to the band  region $\Delta < |E| < |\varepsilon|$.
Replacing $\varepsilon$ with $-\varepsilon$ in Eqs.~(\ref{app-v}), (\ref{app-zeta}), (\ref{app-alpha}), and (\ref{app-beta}), we obtain
 $v_\pm', \zeta_{\uparrow,\pm}, \alpha_\pm',\beta_\pm'$, and
 \newpage
\begin{eqnarray}
S_\uparrow
&=&
\left(
\begin{array}{ccc}
\frac{\alpha_+^{\prime 2}}{v_+'}  &
\frac{-\alpha'_+\alpha'_-}{v_+'}\frac{\beta_+'}{\beta_-'} &
\frac{\alpha_+'}{v_+'}\beta_+'
\\
\frac{-\alpha_+'\alpha_-'}{v_-'} \frac{\beta_-'}{\beta_+'}&
\frac{\alpha_-^{\prime 2}}{v_-'} &
\frac{-\alpha'_-}{v_-'}\beta_-'
\\
\frac{\alpha'_+}{\beta_+'}&
-\frac{\alpha'_-}{\beta_-'}&
1
\\
\end{array}
\right)\otimes \frac{u_+}{\zeta_{\uparrow,+}} 
\nonumber \\
& &
+\left(
\begin{array}{ccc}
\frac{1}{v_+'}  &
\frac{\beta_+'}{v_+'\beta_-'}&
\frac{1}{v_+'}\beta_+'
\\
\frac{\beta_-'}{v_-'\beta_+'} &
\frac{1}{v_-'} &
\frac{1}{v_-'}\beta_-'
\\
\frac{1}{\beta_+'}&
\frac{1}{\beta_-'}&
1
\\
\end{array}
\right)\otimes \frac{u_-}{\zeta_{\uparrow,-}} -\bf{1}_6
\label{app-S-u}
\end{eqnarray}
Useful formulas are $\beta_\pm'=\frac{1}{\beta_\pm}$ and $\beta_+\beta_-=\frac{\varepsilon+E}{\varepsilon-E} =-\frac{\alpha_\pm}{\alpha_\pm'}$.
When $|\varepsilon| > |E|> \Delta$, $v_\pm$  is  proportional to the probability flow of mode $\eta_{\tau,\pm}$, and
 the normalized scattering matrix $\bar{S}_\circ=VS_{\circ}V^{-1}$ becomes
the 6-dimensional unitary matrix, where 
\begin{equation}
V=\left(
\begin{array}{ccc}
 \sqrt{v_+} & 0 & 0 \\
 0 & \sqrt{v_-} &0 \\
  0 & 0& 1 
\end{array}
\right)
\otimes \bf{1}_2.
\end{equation}
When $\Delta>|E|$ and the length of the bilayer region becomes infinite, there is perfect reflection of an incident wave from the monolayer region, and the (3,3) block $\zeta_{\circ,+}^{-1}u_++\zeta_{\circ,-}^{-1}u_--{\bf 1}_2$  becomes the two-dimensional unitary matrix.
Correspondence between Eq.~(\ref{multiple-T}) and $S_\circ$ is as follows.
 $r_\circ$ consists of the (1,1),(1,2), (2,1) and (2,2) blocks of $S_\circ$.
  $t_\downarrow$ includes the (1,3) and (2,3) blocks of $S_\downarrow$.
  $\tilde{t}_\bullet$ is composed of the (3,1) and (3,2) blocks of $S_\bullet$.

In the following, we consider only the gapped region, $\Delta>|E|$.
 Applying Eqs.~(\ref{app-S-d}) and (\ref{app-S-u}) to Eq.~(\ref{direct-T}),
 we obtain
\begin{equation}
T_{\nu',\nu}=\frac{\gamma_1^2}{q^2} \left|\sum_{\tau=\pm} \frac{(E+\varepsilon)Y^{(\tau)}_{\nu',\nu}}{\zeta_{\bullet,+}\zeta_{\bullet,-}\zeta_{\downarrow,+}\zeta_{\downarrow,-}}\right|^2
\label{app-T-Y}
\end{equation} 
where 
\begin{eqnarray}
Y_{\nu',\nu}^{(\tau)}  &=&
\zeta_{\downarrow, -}^2{\rm Re}\left( \frac{\alpha}{\beta}\lambda_{\tau}^N \right) 
+
\nu\nu' \zeta_{\downarrow, +}^2{\rm Re}\left( \frac{\lambda_{\tau}^N}{\alpha\beta}\right) \nonumber \\
&&
+
2\nu\delta_{\nu',\nu} \zeta_{\downarrow, +}\zeta_{\downarrow, -} \tau {\rm Re}\left( \frac{\lambda_\tau^N}{\beta}\right) 
\label{app-Y-DD-1}
\end{eqnarray}
for the $\downarrow$ junction, and
\begin{eqnarray}
Y_{\nu',\nu}^{(\tau)} &= &
 \zeta_{\downarrow, -}\zeta_{\uparrow, -}{\rm Re}\left(\alpha' \lambda_{\tau}^N \right) 
+
\nu\nu' \zeta_{\downarrow, +} \zeta_{\uparrow,+}{\rm Re}\left( \frac{\lambda_{\tau}^N}{\alpha}\right) \nonumber \\
&& 
+\left(\nu \zeta_{\uparrow, -}\zeta_{\downarrow, +}\frac{\alpha'}{\alpha}+\nu'\zeta_{\uparrow, +}\zeta_{\downarrow, -}\right) 
 \tau{\rm Re}\left(\lambda_\tau^N\right) 
\label{app-Y-UD-1}
\end{eqnarray}
for the $\uparrow$ junction with abbreviation $\beta=\beta_+$, $\alpha=\alpha_+$ and $\alpha'=\alpha_+'$.
When $|\varepsilon| \gg \gamma_1$, all the $\zeta$'s are close to $1-2i|\varepsilon|\frac{E}{q}$,  and $(\alpha,\alpha')  \simeq \frac{\varepsilon}{|\varepsilon|}(\alpha_0,\alpha_0')$, where  $\alpha_0\equiv\frac{|E+\varepsilon|}{\sqrt{p}}$,  and 
$\alpha_0' \equiv \frac{\varepsilon}{|\varepsilon|}\frac{E-\varepsilon}{\sqrt{p}}$.
This leads to the following approximations
\ \\
\ \\
\ \\
\begin{widetext}
\begin{equation}
Y_{\nu',\nu}^{(\tau)}= 
%\frac{\gamma_1^4}{q^4} (E+\varepsilon)(\varepsilon^2-E^2) 
\left(1-2i|\varepsilon|\frac{E}{q}\right)^2 
\left[\left(\sqrt{\alpha_0}+\frac{\varepsilon}{|\varepsilon|}\frac{\tau\nu}{\sqrt{\alpha_0}}\right)^2\delta_{\nu',\nu}
+ \left(\alpha_0-\frac{1}{\alpha_0}\right)\delta_{\nu',-\nu}
\right]\frac{\varepsilon}{|\varepsilon|} {\rm Re}\left( \frac{\lambda_\tau^N}{\beta}\right) 
\label{app-Y-DD-2}
\end{equation}
for Eq.~(\ref{app-Y-DD-1}), and 
\begin{equation}
Y_{\nu',\nu}^{(\tau)}= 
%\frac{\gamma_1^4}{q^4} (E+\varepsilon)(\varepsilon^2-E^2) 
\left(1-2i|\varepsilon|\frac{E}{q}\right)^2 
\left[
\left(\alpha_0'
+\frac{1}{\alpha_0}
+\frac{\varepsilon}{|\varepsilon|}\tau\nu\frac{2E}{E+\varepsilon}
\right)\delta_{\nu',\nu}
+
\left(
\alpha_0'
-\frac{1}{\alpha_0}
-\frac{\varepsilon}{|\varepsilon|}\tau\nu\frac{2\varepsilon}{E+\varepsilon}
\right)\delta_{\nu',-\nu}
\right]
\frac{\varepsilon}{|\varepsilon|} {\rm Re}\left(\lambda_\tau^N\right) 
\label{app-Y-UD-2}
\end{equation}
 for Eq.~(\ref{app-Y-UD-1}).
 \end{widetext}
 Since $|E| \ll |\varepsilon|$,  $\alpha_0 >0$, and $\alpha_0' <0$, the terms satisfying Eq.~(\ref{dominant-path}) dominate Eqs.~(\ref{app-Y-DD-2}) and (\ref{app-Y-UD-2}).
 In comparison with Eq.~(\ref{T-smat}), $|Y^{(+)}+Y^{(-)}|$ equals $\frac{\gamma_1^2}{q^2}|\varepsilon - E||Z|$ in these dominant terms.
 
 The substitution of the $\zeta$  with $1-2i|\varepsilon|\frac{E}{q}$ is applied only to $Y$ and not to the denominator
   $\zeta_{\bullet,+}\zeta_{\bullet,-}\zeta_{\downarrow,+}\zeta_{\downarrow,-}$  in Eq.~(\ref{app-T-Y}).
   Had the substitution been performed in the denominator as well, the accuracy of the derived analytical expression would have been compromised. This substitution was used within the minimal necessary extent to reproduce the Re factors in Eqs.~(\ref{xi-j-d}) and (\ref{xi-j-u}).
 Regardless of whether Eqs.~(\ref{app-Y-UD-1})  or (\ref{app-Y-UD-2}) is used, Eq.~(\ref{app-T-Y}) satisfies the symmetry (\ref{UD-E-sym})
 owing to Eq.~(\ref{zeta-E-sym}).


\begin{thebibliography}{99}
\bibitem{note}
When $k_y=0$, we can use the analytical expressions demonstrated in Appendix B. 
For non-zero $k_y$, however, we have to perform numerical calculations.
  It remains challenging to derive a clear analytical expression for $t_\downarrow$ and $\tilde{t}_\bullet$ 
at finite $k_y$.
\end{thebibliography}
\end{document}